\documentclass[graybox]{svmult}

\usepackage{mathptmx}       
\usepackage{helvet}         
\usepackage{courier}        
\usepackage{type1cm}        
\usepackage{graphicx}        
\usepackage[bottom]{footmisc} 

\newcommand{\beq}{\begin{equation}}
\newcommand{\eeq}{\end{equation}}
\newcommand{\bd}{\begin{displaymath}}
\newcommand{\ed}{\end{displaymath}}
 
\newcommand{\p}{\partial}        
\newcommand{\rmd}{{\rm d}}
\newcommand{\Om}{\Omega}
\newcommand{\Omk}{\Omega_{\rm K}}

\newcommand{\aaa}[1]{{\it Astron.\ Astrophys.} {\bf #1}}              
\newcommand{\aap}[1]{{\it Astron.\ Astrophys.} {\bf #1}}              
\newcommand{\araa}[1]{{\it Ann. Rev. Astron.\ Astrophys.} {\bf #1}}

\newcommand{\apj}[1]{{\it Astrophys. J.} {\bf #1}}                    
\newcommand{\apjl}[1]{{\it Astrophys.\ J.\ Letters} {\bf #1}}          
\newcommand{\apjs}[1]{{\it Astrophys.\ J.\ Suppl.} {\bf #1}}           
\newcommand{\ass}[1]{{\it Ap.\ Space Sci.} {\bf #1}}

\newcommand{\mnras}[1]{{\it Mon.\ Not.\ R.\ Astron.\ Soc.} {\bf #1}}   
\newcommand{\nature}[1]{{\it Nature} {\bf #1}}

\newcommand{\pasj}[1]{{\it Publ.\ Astr.\ Soc.\ Japan} {\bf #1}}        
\newcommand{\pasp}[1]{{\it Publ.\ Astr.\ Soc.\ Pac.} {\bf #1}}

\newcommand{\gapprox}{\;\rlap{\lower 2.5pt
             \hbox{$\sim$}}\raise 1.5pt\hbox{$>$}\;}
\newcommand{\lapprox}{\;\rlap{\lower 2.5pt
             \hbox{$\sim$}}\raise 1.5pt\hbox{$<$}\;}
\newcommand{\bfg}[1]{\setbox0=\hbox{#1}%
  \kern-.025em\copy0\kern-\wd0
  \kern.05em\copy0\kern-\wd0
  \kern-.025em\raise.0433em\box0}
  
\newcommand{\rea}{ \noindent \hangindent 10pt}
\begin{document}

  \title*{Accretion Disks}  
  \author{{\bf H.C.\ Spruit} \\
   ~ \\
  Max Planck Institute for Astrophysics, \\
  Box 1317, 85741 Garching, Germany, \\{\tt henk@mpa-garching.mpg.de}}

\authorrunning{H.C.\ Spruit}
\maketitle
\vspace{-1 cm}

\abstract{
In this lecture the basic theory of accretion disks is reviewed, with emphasis on
aspects relevant for X-ray binaries and Cataclysmic Variables. The text gives a 
general introduction as well as a selective discussion of a number of more recent topics.}

\section{Introduction}
Accretion disks are inferred to exist as objects of very different scales: 
millions of km in low Mass X-ray Binaries (LMXB) and Cataclysmic Variables (CV),
solar radius-to-AU scale disks in protostellar objects, and AU-to-parsec scale 
disks in Active Galactic Nuclei (AGN). 

An interesting observational connection exists between accretion disks and 
jets (such as the spectacular jets from AGN and protostars), and outflows 
(the `CO-outflows' from protostars and the `broad-line-regions' in 
AGN).  Lacking direct (i.e. spatially resolved) observations of disks, theory has
tried to provide models, with varying degrees of success. Uncertainty still exists
with respect to some basic questions. In this situation, progress
made by observations or modeling of a particular class of objects has 
direct impact for the understanding of other objects, including the enigmatic
connection with jets.

In this lecture I concentrate on the more basic aspects of accretion disks, but an
attempt is made to mention topics of current interest as well. Some emphasis is 
on those aspects of  accretion disk theory that connect to the observations 
of LMXB and CV's. 
For other reviews on the basics of accretion disks, see Pringle (1981), Papaloizou 
and Lin (1995).  For a more extensive introduction, the textbook by Frank et al. 
(2002). For a comprehensive text on CVs, Warner 1995. 

\section{Accretion: general}

\label{bondi}
Gas falling into a point mass potential
\bd \Phi=-{GM\over r} \ed
from a distance $r_0$ to a distance $r$ converts gravitational into kinetic
energy, by an amount $\Delta\Phi=GM(1/r-1/r_0)$. For simplicity, assuming that the
starting distance is large, $\Delta\Phi=GM/r$. The speed of arrival, the 
{\em free-fall speed} $v_{\rm ff}$ is given by 
\beq {1\over 2}v_{\rm ff}^2=GM/r.\eeq
If the gas is then brought to rest, for example at the surface of a star, the amount 
of energy $e$ dissipated per unit mass is  
\bd e={1\over 2}v_{\rm ff}^2={GM\over r} \qquad (\mathrm rest). \ed
If, instead, it goes into a circular Kepler orbit at distance $r$:
\bd e={1\over 2}{GM\over r} \qquad (\mathrm orbit). \ed
The dissipated energy may go into internal energy of the gas, and into radiation
which escapes to infinity (usually in the form of photons, but neutrino losses
can also play a role in some cases).

\subsection{Adiabatic accretion}
\label{tvir}
Consider first the case when radiation losses are neglected. Any mechanical energy 
dissipated stays locally in the flow. This is called an {\it adiabatic} flow (not to be
confused with {\em isentropic} flow). For an ideal gas with constant ratio of specific 
heats $\gamma$, the internal energy per unit mass is
\bd e={P\over (\gamma -1)\rho}. \ed
With the equation of state
\beq P={\cal R}\rho T/\mu \label{es}\eeq
where ${\cal R}$ is the gas constant and $\mu$ the mean atomic weight per particle,
we find the temperature of the gas after the dissipation has taken place (assuming
that the gas goes into a circular orbit): 
\beq T={1\over 2}(\gamma-1)T_{\rm vir}, \label{ttv}\eeq
where $T_{\rm vir}$, the {\it virial temperature} is defined as
\bd T_{\rm vir}={GM\mu\over{\cal R}r}=g\,r\mu/{\cal R}, \ed
where $g$ is the acceleration of gravity at distance $r$.
In an atmosphere with temperature near $T_{\rm vir}$, the sound speed \-
$c_{\rm s}=(\gamma{\cal R}T/\mu)^{1/2}$ is close to the escape speed from 
the system, and the hydrostatic pressure scale height, $H\equiv {\cal R}T/(\mu g)$ 
is of the order of $r$. Such an atmosphere may evaporate on a relatively short
time scale in the form of a stellar wind. This is as expected from energy 
conservation: if no energy is lost through radiation, the energy gained 
by the fluid while falling into a gravitational potential is also sufficient to move 
it back out again. 

A simple example is spherically symmetric adiabatic accretion (Bondi, 1952). 
An important result is that such accretion is possible only if $\gamma\le 5/3$. 
The lower $\gamma$, the lower the temperature in the accreted gas 
(eq.~\ref{ttv}), and the easier it is for the gas to stay bound in the potential. 
A classical situation where adiabatic and roughly spherical accretion takes
place is a supernova implosion: when the central temperature becomes high
enough for the radiation field to start disintegrating nuclei, $\gamma$ drops
and the envelope collapses onto the forming neutron star via an
accretion shock. Another case are Thorne-Zytkow objects (e.g. Cannon et al. 1992),
where $\gamma$ can drop to low values due to pair creation,
initiating an adiabatic accretion onto the black hole.

Adiabatic spherical accretion is fast, taking place on the dynamical time scale: 
something on the order of the free fall time scale, or Kepler orbital time scale,
\beq \tau_{\rm d}=r/v_{\rm K}=\Omega_{\rm K}^{-1}=(r^3/GM)^{1/2}, \eeq
where $v_{\rm K},\,\Omega_{\rm K}$ are the Kepler orbital velocity and angular 
frequency.

When radiative loss becomes important, the accreting gas can stay cool
irrespective of the value of $\gamma$, and Bondi's critical value $\gamma=5/3$
plays no role. With such losses, the temperatures of accretion disks are usually
much lower than the virial temperature.  

\subsection{Temperature near compact objects}
For accretion onto a neutron star surface, $R=10$ km, $M=1.4M_\odot$, we have 
a free fall speed $v_{\rm ff}/c\approx 0.4\, c$ (this in Newtonian approximation, 
the correct value in General Relativity is quantitatively somewhat different). The
corresponding virial temperature would be $T_{\rm v}\sim 2~10^{12}$ K, equivalent 
to an average energy of 150 MeV per particle.  

This is not the actual temperature we 
should expect, since other things happen before such temperatures are reached. If
the accretion is adiabatic, one of these is the creation of a very dense radiation 
field. The energy liberated per infalling particle is still the same, but it gets shared 
with a large number of photons. At temperatures above the electron rest mass 
($\approx 0.5 MeV$), electron-positron pairs $e^\pm$ are produced in addition 
to photons. These can take up most of the accretion energy, limiting the temperature 
typically to a few MeV. 

In most observed disks temperatures do not get even close to an MeV, however, 
because accretion is rarely adiabatic. Energy loss takes place by escaping photons 
(or under more extreme conditions: neutrinos). Exceptions are the
radiatively inefficient accretion flows discussed in section \ref{isaf}.

\subsubsection{Radiative loss}
Next to the adiabatic temperature estimates, a useful characteristic number is the 
`black body effective temperature'. Here, the approximation made is that the accretion
energy is radiated away from an optically thick surface of some geometry, under the 
assumption of a balance between the heating rate by release of accretion energy and 
cooling by radiation. For a specific example consider the surface of a star of radius 
$R$ and mass $M$ accreting via a disk. Most of the gravitational energy is released 
close to the star, in a region with surface area of the order, let's call it $4\pi R^2$. 
If the surface radiates approximately as a black body of temperature $T$ (make a note 
of the fact that this is a bad approximation if the opacity is dominated by electron 
scattering), the balance would be
\beq \dot M{G M\over R}\approx 4\pi R^2\sigma_{\rm r} T^4, \eeq
where $\sigma_{\rm r}$ is the Stefan-Boltzmann constant, $\sigma_{\rm r}=a_{\rm r}c/4$ 
if $a_{\rm r}$ is Planck's radiation constant.
For a neutron star with $M=1.4 M_\odot=3\,10^{33}$ g, $R=10$ km, accreting at a
typical observed rate (near Eddington rate, see below), $\dot M= 10^{18}$g/s 
$\approx 10^{-8}M_\odot$/yr, this temperature would be $T\approx 10^7$ K, or 
$\approx 1$ keV per particle. Radiation with this characteristic temperature is 
observed in accreting neutron stars and black holes in their so-called `soft X-ray states'
(as opposed to their `hard' states, in which the spectrum is very far from a black body). 

The kinds of processes involved in radiation from accretion flows form a large 
subject in itself and will not be covered here. For introductions see Rybicki \& 
Lightman (1979) and Frank et al. (2002). At the moderate temperatures encountered
in protostars and white dwarf accreters, the dominant processes are the ones
known from stellar physics: molecular and atomic transitions and Thomson scattering. 
Up to photon energies around 10 keV (the Lyman edge of an iron nucleus with one
electron left) these processes also dominate the spectra of neutron star and black 
hole accreters. Above this energy, observed spectra become dominated by Compton 
scattering. Cyclotron/synchrotron radiation plays a role in when a strong magnetic
field is present, which can happen in most classes of accreting objects.

\subsection{Critical luminosity}
\label{eddi}
Objects of high luminosity have a tendency to blow their atmospheres away due
to the radiative force exerted when the outward traveling photons are scattered
or absorbed. Consider a volume of gas on which a flux of photons is incident
from one side. Per gram of matter, the gas presents a scattering (or
absorbing) surface area of $\kappa$ cm$^2$ to the escaping radiation. The force 
exerted by the radiative flux $F$ on one gram is $F\kappa/c$. The force of 
gravity pulling back on this one gram of mass is $GM/r^2$. The critical flux  at 
which the two forces balance (energy per unit area and time) is
\beq F_{\rm E}={c\over\kappa}{GM\over r^2}. \label{fe}\eeq
Assuming that this flux is {\em spherically symmetric}, it can be converted into a
luminosity,
\beq L_{\rm E}=4\pi GMc/\kappa ,\label{le}\eeq
the Eddington critical luminosity, popularly called the Eddington limit 
(e.g. Rybicki and Lightman, 1979). If the gas is fully
ionized, its opacity is dominated by electron scattering, and for solar
composition $\kappa$ is then of the order $0.3$ cm$^2$/g (about a factor 2
lower for fully ionized helium). With these assumptions, 
\bd L_{\rm E}\approx 1.7\,10^{38}{M\over M_\odot}~~{\mathrm erg/s}\approx
4\,10^4{M\over M_\odot}~L_\odot .\label{Ledd}\ed
This number is different if the opacity is not dominated by electron scattering. In 
partially ionized gases of solar composition bound-bound and bound-free 
transitions can increase the opacity by a factor up to $10^3$; the Eddington flux
is then correspondingly lower.

If the luminosity results from accretion, one can define a corresponding Eddington 
characteristic accretion rate $\dot M_{\rm E}$:
\beq {GM\over r}\dot M_{\rm E}=L_{\rm E}\quad\rightarrow\quad\dot M_{\rm E}=4\pi
rc/\kappa \label{me}.\eeq
With $\kappa=0.3$:
\bd \dot M_{\rm E}\approx 1.3\,10^{18}r_6~~{\mathrm g/s}\approx 2\,10^{-8}
r_6~~M_\odot{\mathrm yr}^{-1}, \ed
where $r_6$ is the radius of the accreting object in units of $10^6$ cm. The 
characteristic accretion rate thus scales with the {\em size} of the accreting object,
while the critical luminosity scales with {\em mass}.

Whereas $L_{\rm E}$ is a critical value which in several circumstances plays 
the role of a limit, the Eddington characteristic accretion rate is less of a limit.  
For more on exceptions to $L_{\rm E}$ and $\dot M_{\rm E}$ see \ref{limitsE} below.

\subsubsection{Eddington luminosity at high optical depth}
The Eddington characteristic luminosity was derived above under the assumption
of a radiation flux passing through an optically thin medium surrounding the 
radiation source. What changes if the radiation passes through an optically thick
medium, such as a stellar interior? At high optical depth the radiation field 
can be assumed to be nearly isotropic, and the {\em diffusion approximation} 
applies (cf. Rybicki \& Lightman 1979). The radiative heat flux can then be written
in terms of the radiation pressure $P_{\rm r}$ as 
\beq  F_{\rm r}=c{\rmd P_{\rm r}\over\rmd\tau},\label{frad}\eeq
where $\tau$ is the optical depth
\beq \rmd\tau=\kappa\rho\,\rmd s\eeq
along a path $s$, and $\kappa$ an appropriate frequency-averaged opacity (such as 
the Rosseland mean). Balancing the gradient of the radiation pressure against the 
force of gravity gives the maximum radiation pressure that can be supported:
\beq {\bf \nabla}P_{\rm r, max}={\bf g}\rho,\eeq
where $\bf g$ is the acceleration of gravity. With (\ref{frad}) this yields the maximum 
radiation flux at a given point in a static gravitating object:
\beq {\bf F}_{\rm r,max}= c\,{\bf g}/\kappa,\label{FE}\eeq
i.e. the same as the critical flux in the optically thin case. 

\subsubsection{Limitations of the Eddington limit} 
\label{limitsE}
The derivation of $F_{\rm E}$ assumed that the force relevant in the argument is
gravity. Other forces can be larger. An example would be 
a neutron star with a strong magnetic field. The curvature force $B^2/(4\pi r_{\rm c})$ 
in a loop of magnetic field (where $r_{\rm c}$ is the radius of curvature of the field 
lines) can balance a pressure gradient $\sim P/r_{\rm c}$, so the maximum pressure
that can be contained in a magnetic field\footnote{Depending on circumstances the 
actual maximum is less than this because a magnetically contained plasma tends to
 `leak across' field lines through MHD instabilities.}  is of order $P\sim B^2/{8\pi}$. If the 
pressure is due to radiation, assuming an optical depth $\tau\ge 1$ so the diffusion 
approximation can be used,  the maximum radiative energy flux is then of the order
\beq 
F_{\rm r,max} \approx c P_{\rm r}/\tau\approx c{B^2\over 8\pi\tau}.
\eeq
In the range of validity of the assumptions made this has its maximum for an 
optical depth of order 
unity:  $F_{\rm r,max}\approx c{B^2/ 8\pi}$. For a neutron star of radius $R=10^ 6$ cm 
and a field strength of $10^{12}$ G, this gives $L_{\rm r,max} \approx 10^{46}$ erg/s,  
many orders of magnitude higher than the Eddington value $L_{\rm E}$.
[This explains the enormous luminosities that can be reached in so-called {\em 
magnetar outbursts}, e.g.\ Hurley et al. 2005].

The Eddington argument considers only the radiative flux. Larger energy fluxes 
are possible if energy is transported by other means, for example by convection. 

Since $L_{\rm E}$ depends on opacity, it can happen that $L_{\rm E}$ is 
lower in the atmosphere of a star than in its interior. A luminous star radiating near 
its (internal) Eddington rate will then blow off its atmosphere in a {\em radiatively 
driven stellar wind}; this happens for example in Wolf-Rayet stars.
In the context of protostellar accretion, the opacity  in the star forming cloud from 
which the protostar accretes is high due to atomic and molecular transitions. As 
a result, the radiation pressure from a massive (proto-)star, with a luminosity 
approaching (\ref{Ledd}), is able to clear the away the molecular cloud from which 
it formed. This is believed to set a limit on the mass that can be reached by
a star formed in a molecular cloud.

\subsubsection{Neutron stars vs. black hole accreters}
In deriving the critical accretion rate, it was assumed that the gravitational energy 
liberated is emitted in the form of radiation. In the case of a black hole accreter,
this is not necessary since mass can flow through the hole's horizon, taking with it all 
energy contained in it. Instead of being emitted as radiation, the energy adds to the 
mass of the hole. This becomes especially important at high accretion rates, 
$\dot M>\dot M_{\rm E}$ (and in the ion supported accretion flows discussed in 
section \ref{isaf}). The parts of the flow close to the hole then become optically thick, 
the radiation stays trapped in the flow, and instead of producing luminosity gets 
swallowed by the hole. 
The accretion rate on a black hole can thus be arbitrarily large in principle (see 
also section \ref{radv}, and chapter 10 in Frank et al, 2002). 

A neutron star cannot absorb this much energy (only a negligible amount is taken 
up by conduction of heat into its interior), so $\dot M_{\rm E}$ is more relevant for 
neutron stars than for black holes. It is not clear to what extent it actually limits the 
possible accretion rate, however, since the limit was derived under the assumption of 
spherical symmetry. It is possible that accretion takes place in the form of an optically 
thick disk, while the energy released at the surface produces an outflow along the 
axis, increasing the maximum possible accretion rate (cf. discussion in  \ref{radv}). 
This has been proposed (e.g. King, 2004) as a possible conservative interpretation 
of the so-called ultraluminous X-ray sources (ULX), rare objects with luminosities of 
$10^{39}-10^{41}$  erg/s. These are alternatively and more excitingly  suggested to 
harbor intermediate mass black holes (above $\approx 30 M_\odot$).

\section{Accretion with Angular Momentum}
When the accreting gas has a zonzero angular momentum with respect to the
accreting object, it can not accrete directly. A new time scale then plays a role,
the time scale for outward transport of angular momentum. Since this is in
general much longer than the dynamical time scale, much of what was said about
spherical accretion needs modification for accretion with angular momentum.

Consider the accretion in a close binary consisting of a compact (white dwarf,
neutron star or black hole) primary of mass $M_1$ and a main sequence
companion of mass $M_2$. The mass ratio is defined as $q=M_2/M_1$ (note:
in the literature $q$ is just as often defined the other way around).

If  $M_1$ and $M_2$ orbit each other in a circular orbit and their separation
is $a$, the orbital frequency $\Omega$ is
\bd \Omega^2=G(M_1+M_2)/a^3 .\ed
The accretion process is most easily described in a coordinate frame that
corotates with this orbit, and with its origin in the center of mass. Matter
that is stationary in this frame experiences an effective potential, the
{\it Roche potential} (Ch. 4 in Frank, King and Raine, 2002), given by
\beq 
\phi_{\rm R}({\mathbf r})=-{GM\over r_1}-{GM\over r_2}-{1\over 2}
\Omega^2\varpi^2 \label{ro}
\eeq 
where $r_{1,2}$ are the distances of point $\mathbf r$ to stars 1,2, and $\varpi$ 
the distance from the rotation axis (the axis through the center of mass, 
perpendicular to the orbit ). Matter that
does {\it not} corotate experiences a very different force (due to the Coriolis
force). The Roche potential is therefore useful only in a rather limited
sense. For non-\-corotating gas intuition based on the Roche geometry can be
misleading. Keeping in mind this limitation, consider the equipotential
surfaces of (\ref{ro}). The surfaces of stars $M_{1,2}$, assumed to corotate
with the orbit, are equipotential surfaces of (\ref{ro}). Near the centers of
mass (at low values of $\phi_{\rm R}$) they are unaffected by the other star, 
at higher $\Phi$ they are distorted and at a critical value $\Phi_1$ the two
parts of the surface touch. This is the critical Roche surface $S_1$ whose two
parts are called the Roche lobes. 

Binaries lose angular momentum through
gravitational radiation and a magnetic wind from the secondary (if it has
a convective envelope). Through this loss the separation between the
components decreases and both Roche lobes decrease in size. Mass transfer
starts when $M_2$ fills its Roche lobe, and continues as long as the angular
momentum loss from the system lasts. Mass transfer can also be due to 
expansion of the secondary in course of its evolution, and mass transfer can be
a runaway process, depending on mass and internal structure of the secondary.
This is a classical subject in the theory of binary stars, for an introduction 
see Warner (1995).

\begin{figure}
\hfil \includegraphics[width=10 cm, clip]{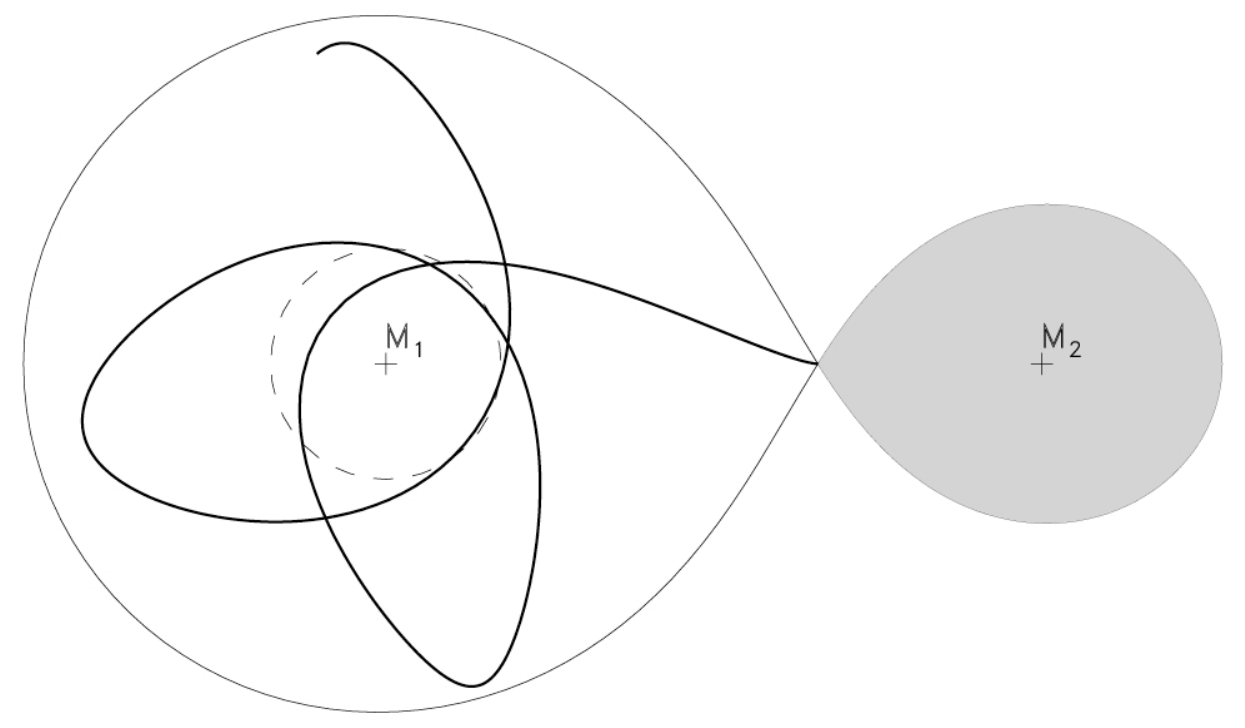}\hfil
 \caption[]{\label{rof}
Roche geometry for $q=0.2$, with free particle orbit from $L_1$ (as seen in a
frame corotating with the orbit). Dashed: circularization radius.}   
\end{figure}

A stream of gas then flows through the point
of contact of the two parts of $S_1$, the inner Lagrange point $L_1$. If the
force acting on it were derivable entirely from (\ref{ro}) the gas would just
fall in radially onto $M_1$. As soon as it moves however, it does not corotate
any more and its orbit under the influence of the Coriolis force is different
(Fig.~\ref{rof}).

Since the gas at $L_1$ is usually very cold compared with the virial temperature, 
 the velocity it acquires exceeds the sound speed already after moving a
 small distance from $L_1$. The flow into the Roche lobe of $M_1$ 
is therefore highly {\it supersonic}. Such hypersonic flow is essentially
ballistic, that is, the stream flows approximately along the path taken by freely 
falling particles. 

Though the gas stream on the whole follows an path close to that of a free
particle, a strong shock develops at the point where the path intersects
itself\footnote{In practice shocks already develop shortly after passing the
pericenter at $M_1$, when the gas is decelerated again. Supersonic flows that
are decelerated, by whatever means, in general develop shocks (e.g. Courant and
Friedrichs 1948, Massey, 1968). The effect can be seen in action in the movie
published in R\'o\.zyczka and Spruit, 1993.}. After this, the gas settles into
a ring, into which the stream continues to feed mass. If the mass ratio $q$ is
not too small this ring forms fairly close to $M_1$. An approximate value for
its radius is found by noting that near $M_1$ the tidal force due to the
secondary is small, so that the angular momentum of the gas with respect to
$M_1$ is approximately conserved. If the gas continues to conserve angular
momentum while dissipating energy, it settles into the minimum energy orbit
with the specific angular momentum $j$ of the incoming stream. The value of 
$j$ is found by a simple integration of the orbit starting at $L_1$ and 
measuring $j$ at some point near pericenter. The radius of the orbit, the 
nominal {\it circularization radius} $r_{\rm c}$ is then defined through  
$ (GM_1r_{\rm c})^{1/2}=j. $ In units of the
orbital separation $a$, $r_{\rm c}$ and the distance $r_{{\rm L}1}$ from $M_1$ to
$L_1$ are functions of the mass ratio only. As an example for $q=0.2$,
$r_{{\rm L}1}\approx 0.66a$ and the circularization radius $r_{\rm c}\approx
0.16a$. In practice the ring forms somewhat outside $r_{\rm c}$, because there is
some angular momentum redistribution in the shocks that form at the impact of the
stream on the ring. The evolution of the ring depends critically on nature and 
strength of the angular momentum transport processes. If sufficient `viscosity' is 
present it spreads inward and outward to form a disk.

At the point of impact of the stream on the disk the energy dissipated is a
significant fraction of the orbital kinetic energy, hence the gas heats up to a
significant fraction of the virial temperature. For a typical system with $M_1=1
M_\odot$, $M_2=0.2 M_\odot$ and an orbital period of 2 hrs, the observed size
of the disk (e.g.\ Wood et al. 1989b, Rutten et al. 1992) $r_{\rm d}/a\approx 0.3$,
the orbital velocity at $r_{\rm d}$ about 900 km/s, and the virial temperature at
$r_{\rm d}$ is $\approx 10^8$ K. The actual temperatures at the impact point are much
lower, due to rapid cooling of the shocked gas. Nevertheless the impact gives rise
to a prominent `hot spot' in many systems, and an overall heating of the outermost
part of the disk.

\section{Thin disks: properties}
\subsection{Flow in a cool disk is supersonic}
Ignoring viscosity, the equation of motion in the potential of a point mass is
\beq 
{\p{\mathbf v}\over\p t}+{\mathbf v}\cdot\nabla{\mathbf v}= -{1\over\rho}\nabla
P-{GM\over r^2}{\mathbf{\hat r}}, 
\eeq
where $\mathbf{\hat r}$ is a unit vector in the spherical radial direction $r$.
To compare the order of magnitude of the terms, choose a position $r_0$ in the
disk, and choose as typical time and velocity scales the orbital time scale
$\Omega_0^{-1}=(r_0^3/GM)^{1/2}$ and velocity $\Omega_0r_0$. For simplicity 
of teh argument, assume a fixed temperature $T$. The pressure gradient term is then
\bd {1\over\rho}\nabla P={{\cal R}\over\mu}T\nabla\ln P. \ed
In terms of the dimensionless quantities 
\bd \tilde r=r/r_0, \qquad \tilde v=v/(\Omega_0r_0), \ed 
\bd \tilde t=\Omega_0t, \qquad \tilde\nabla=r_0\nabla, \ed
the equation of motion becomes
\beq {\p{\mathbf {\tilde v}}\over\p \tilde t}+{\mathbf {\tilde v}}\cdot
\tilde\nabla 
{\mathbf{\tilde v}}=-{T\over T_{\rm vir}}\tilde\nabla\ln P-{1\over\tilde r^2}
{\mathbf{\hat r}}.
\eeq
All terms and quantities in this equation are of order unity by the assumptions
made, except the pressure gradient term which has the coefficient $T/T_{\rm vir}$ 
in front. If cooling is important, so that $T/ T_{\rm vir}\ll 1$, the pressure term is
negligible to first approximation. Equivalent statements are also
that the gas moves hypersonically on nearly Keplerian orbits, and that the disk is
thin, as is shown next.

\subsection{Disk thickness}
\label{height}
The thickness of the disk is found by considering its equilibrium in the
direction perpendicular to the disk plane. In an axisymmetric disk, using
cylindrical coordinates ($\varpi,\phi,z$), consider the forces at a point ${\mathbf
r}_0$ ($\varpi,\phi,0$) in the midplane, in a frame rotating with the Kepler rate
$\Omega_0$ at that point. The gravitational acceleration
$-GM/r^2{\mathbf{\hat r}}$ balances the centrifugal acceleration 
$\Omega_0^2$\,\bfg{$\hat\varpi$}\  at this point, but not at some distance $z$ above it 
because gravity and centrifugal acceleration work in different directions. Expanding both
accelerations near ${\mathbf r}_0$, one finds a residual acceleration toward the
midplane of magnitude \bd g_z=-\Omega_0^2z. \ed Assuming an isothermal 
gas at temperature $T$, the condition for equilibrium in the $z$-
direction under this acceleration yields a hydrostatic density distribution
\bd \rho=\rho_0(\varpi)\exp\left(-{z^2\over 2H^2}\right). \ed
$H(\varpi)$, called the {\it scale height} of the disk or simply the disk thickness, 
is given in terms of the isothermal sound speed $c_{\rm i}=({\cal R}T/\mu)^{1/2}$ by
\beq H=c_{\rm i}/\Omega_0. \label{h}\eeq
Define $\delta\equiv H/r$, the {\it aspect ratio} of the disk;
it can be expressed in several equivalent ways: 
\beq
\delta={H\over r}={c_{\rm i}\over \Omega r}={1\over \rm Ma}=\left ({T\over T_{\rm vir}}
\right) ^{1/2}, \label{delta}\eeq
where $\mathrm Ma$ is the Mach number of the orbital motion. 

\subsection{Viscous spreading}
\label{spread}
The shear flow between neighboring Kepler orbits in the disk causes friction if some
form of viscosity is present. The frictional torque is equivalent to exchange of angular 
momentum between these orbits. But since the orbits are close to Keplerian, a change in
angular momentum of a ring of gas also means it must change its distance from the
central mass. If the angular momentum is increased, the ring moves to a larger
radius. In a thin disk angular momentum transport (more precisely a nonzero
divergence of the angular momentum flux) therefore automatically implies
redistribution of mass in the disk.

A simple example (L\"ust 1952, see also Lynden-Bell and Pringle 1974) is a narrow
ring of gas at some distance $r_0$. If at $t=0$ this ring is released to evolve under the 
viscous torques, one finds that it first spreads into an asymmetric hump. The hump 
quickly spreads inward onto the central object while a long tail spreads slowly outwards 
to large distances. As $t\rightarrow\infty$ almost all the {\it mass} of the ring accretes 
onto the center, while a vanishingly small fraction of the gas carries almost all the 
{\it angular momentum} to infinity. 

As a result of this asymmetric behavior essentially all the mass of a disk can accrete 
even if the total angular momentum of the disk is conserved. In practice, however, 
there is often an external torque removing angular momentum from the disk: when
the disk results from mass transfer in a binary system. The tidal forces exerted by the 
companion star take up angular momentum from the outer parts of the disk, limiting its 
outward spreading.

\subsection{Observational evidence of disk viscosity}
Evidence for the strength of the angular momentum transport processes in
disks comes from observations of variability time scales. This evidence is
not good enough to determine whether the processes really behave in the same 
way as viscosity, but if this is assumed, estimates can be made
of its magnitude.

Observations of Cataclysmic Variables (CV) give the most detailed information. These 
are binaries with white dwarf (WD) primaries and (usually) main sequence companions 
(for reviews see Meyer-Hofmeister and Ritter 1993,  1995, Warner 1995). A 
subclass of these systems, the Dwarf Novae, show semiregular outbursts. In the 
currently most developed theory, these outbursts are due to an instability in the disk 
(Smak 1971, Meyer and Meyer-Hofmeister 1981, King 1995, Hameury et al. 1998). 
The outbursts are episodes of enhanced  accretion of mass from the
disk onto the primary, involving a significant part of the whole disk. The decay
time of the burst is thus a measure of the viscous time scale of the disk (the
quantitative details depend on the model, see Cannizzo et al. 1988, Hameury et al.
1998): \bd t_{\rm visc}=r_{\rm d}^2/\nu ,\ed where $r_{\rm d}$ is the size of the
disk, $\sim 10^{10}$ cm for a CV. With decay times on the order of days, this yields 
viscosities of the order $10^{15}$ cm$^2$/s, some 14 orders of magnitude above 
the microscopic (`molecular') viscosity of the gas.

Other evidence comes from the inferred time scale on which disks around
protostars disappear, which is of the order of $10^7$ years (e.g.\ Strom et al.,
1993).

\subsection{$\alpha$-parameterization}
Several processes have been proposed to account for these short time scales 
and the large apparent viscosities inferred from them. One of these is that accretion 
does not in fact take place through a viscous-like process as described above at all, 
but results from angular momentum loss through a magnetic wind driven from the 
disk surface (Bisnovatyi-Kogan \& Ruzmaikin 1976, Blandford 1976), much in the 
way sun-like stars spin down by angular momentum loss through their stellar winds. 
The extent to which this plays a role in accretion disks is still uncertain. It would be 
a `quiet' form of accretion, since it can do without energy dissipating processes like 
viscosity. It has been proposed as the explanation for the low ratio of X-ray luminosity 
to jet power in many radio sources (see Migliari and Fender 2006 and references 
therein). This low ratio, however, is also 
plausibly attributed to the low efficiency with which X-rays are produced in the `ion
 supported accretion flows'  (see section \ref{isaf} below) which are expected to be the 
 source of the jet outflows observed from X-ray binaries and AGN. 

The most quantitatively developed mechanism for angular momentum 
transport is a form of magnetic viscosity (anticipated already in Shakura 
and Sunyaev 1973). This is discussed below in section \ref{magdisk}. It requires 
the accreting plasma to be sufficiently electrically conducting. This is often 
the case, but not always: it is questionable for example in the cool outer parts 
of protostellar disks. Other mechanisms thus still play a role in the discussion, 
for example spiral shocks (Spruit et al. 1987) and self-gravitating instabilities 
(Paczy\'nski 1978, Gammie 1997).

In order to compare the viscosities of disks in objects with (widely) different sizes
and physical conditions, introduce a dimensionless viscosity $\alpha$: \beq
\nu=\alpha{c_{\rm i}^2\over\Omega}, \label{alf}\eeq where $c_{\rm i}$ is the
isothermal sound speed as before. The quantity $\alpha$ was introduced by 
Shakura and Sunyaev (1973) as a way of parametrizing our ignorance of the angular
momentum transport process, in a way that allows comparison between systems 
of very different size and physical origin (Their definition of $\alpha$ differs a bit, 
by a constant factor of order unity).

\subsection{Causality limit on turbulent viscosity}
How large can the value of $\alpha$ be, on theoretical grounds? As a simple
model, let's assume that the shear flow between Kepler orbits is unstable to
the same kind of shear instabilities found for flows in tubes, channels, near
walls and in jets. These instabilities occur so ubiquitously that the fluid
mechanics community considers them an automatic consequence of a 
high Reynolds number:
\bd {\mathrm Re}={LV\over\nu} \ed
where $L$ and $V$ are characteristic length and velocity scales of the flow.
If this number exceeds about 1000 (for some forms of instability much less),
instability and turbulence are generally observed. It has been argued
(e.g. Zel'dovich 1981) that for this reason hydrodynamic turbulence is the cause
of disk viscosity. Let's look at the consequences of this assumption. If an eddy
of radial length scale $l$ develops due to shear instability, it will rotate at a
rate given by the rate of shear in the flow, $\sigma$, in a Keplerian disk 
\bd \sigma\approx r{\p\Omega\over\p r}=-{3\over 2}\Omega. \ed
The velocity amplitude of the eddy is $V=\sigma l$, and a field of such 
eddies would produce a turbulent viscosity of the order (leaving out numerical 
factors of order unity)
\beq \nu_{\rm turb}\approx l^2\Omega  .\eeq
In compressible flows, there is a maximum to the size of the eddy set by
causality considerations. The force that allows an instability to form a coherently
overturning eddy is the pressure, which transports information about the flow
at the speed  of sound. The eddies formed by a shear instability can therefore
not rotate faster than $c_{\rm i}$, hence their size does not exceed
$c_{\rm i}/\sigma\approx H$ (eq. \ref{delta}). At the same time, the largest 
eddies formed also have the largest contribution to the exchange of angular 
momentum. Thus we should expect that the turbulent viscosity is given by 
eddies with size of the order $H$:
\bd \nu < H^2\Omega, \ed
or 
\bd \alpha < 1.\ed
The dimensionless number $\alpha$ can thus be interpreted as the effective 
viscosity in units of the maximum value expected if a disk were hydrodynamically 
turbulent. 

\subsubsection{Large scale vortices?}
The small size of hydrodynamic eddies expected, $L<H=c_{\rm i}/\Omega$, 
is due to the high Mach number of the flow in an accretion disk; this makes 
a disk behave like a very {\em compressible} fluid. Attempts to construct large 
scale vortices $L\gg H$ in disks using {\em incompressible} fluid analogies 
continue to be made, both analytically and experimentally. Expansions in disk 
thickness as a small parameter then suggest themselves, in analogy with large scale 
flows such as weather systems and tropical storms in the Earth's atmosphere. 
The size of these systems is large compared with the height of the Earth's atmosphere, 
and expansions making use of this can be effective. This tempting analogy 
is misleading for accretion disks, however, since in contrast with the atmosphere
all flows with a horizontal scale exceeding the vertical thickness are supersonic,
and the use of incompressible fluid models meaningless.

\subsection{Hydrodynamic turbulence?}
\label{turbu}
Does hydrodynamical turbulence along these lines actually exist in disks? In the 
astrophysical
community, a consensus has developed that it does {\em not}, certainly not along the 
simple lines suggested by a Reynolds' number argument. An intuitive argument is 
that the flow in a cool disk is close to Kepler orbits, and these are quite stable. 
Numerical simulations of disk-like flows with different rotation profiles do not show 
the expected shear flow instabilities in cases where the angular momentum increases 
outward, while turbulence put in by hand as initial condition decays in such 
simulations (Hawley et al. 1999). 

Analytical work on the problem has not been able 
to demonstrate instability in this case either (for recent work and references see Lesur 
\& Longaretti 2005, Rincon et al. 2007). Existing proposals for disk turbulence involve 
ad-hoc assumptions about the existence of hydrodynamic instabilities (e.g.\ Dubrulle 
1992). However, a subtle form of hydrodynamic angular momentum transport has 
been identified more recently (Lesur \& Papaloizou 2010); it depends on vertical 
stratification of the disk as well as the presence of a convectively unstable r{\em adial} 
gradient.

A laboratory 
analogy is the rotating Couette flow, an experiment with water between differentially
rotating  cylinders. Recent such experiments have demonstrated that turbulence is 
absent in cases where angular momentum increases outward (as in Keplerian rotation)  
at Reynolds' numbers as high as $10^6$ (the Princeton Couette experiment, Ji et al. 
2006).

In view of these negative results, the popular mechanism for angular momentum 
transport in accretion disk has become magnetic: `MRI turbulence' (section 
\ref{magdisk}).  But beware: in the fluid mechanics community it is considered 
crackpot to suggest that at Reynolds numbers like our $10^{14}$ a flow (no matter 
which or where) can be anything but `fiercely turbulent'. If your career depends on 
being friends with this community, it would be wise to avoid discussions about 
accretion disks or Couette experiments.

\section{Thin Disks: equations}
Consider a thin (= cool, nearly Keplerian, cf. section \ref{height}) disk, axisymmetric
but not stationary. Using cylindrical coordinates ($r,\phi,z$), (note that we
have changed notation from $\varpi$ to $r$ compared with  \ref{height}) we define
the {\it surface density} $\Sigma$ of the disk as  
\beq \Sigma=\int_{-\infty}^\infty\rho{\mathrm d}z\approx
2H_0\rho_0, \eeq  
where $\rho_0$, $H_0$ are the density and scaleheight at the midplane. The
approximate sign is used to indicate that the coefficient in front of $H$ in
the last expression actually depends on details of the vertical structure of
the disk. Conservation of mass, in terms of $\Sigma$ is described by 
\beq{\p\over\p t}(r\Sigma)+{\p\over\p r}(r\Sigma v_r)=0. \label{co}\eeq
(derived by integrating the continuity equation over $z$). Since the disk is
axisymmetric and nearly Keplerian, the radial equation of motion reduces to
\beq v_\phi^2={GM/ r}. \eeq
The $\phi$-equation of motion is
\beq
{\p v_\phi\over\p t}+v_r{\p v_\phi\over\p r}+{v_rv_\phi\over
r}=F_\phi,\label{vphi}
\eeq
where $F_\phi$ is the azimuthal component of the viscous force.
By integrating this over height $z$ and using (\ref{co}),
one gets an equation for the angular momentum balance:
\beq 
{\p\over\p t}(r\Sigma\Omega r^2)+{\p\over\p r}(r\Sigma v_r\Omega r^2)=
{\p\over\p r}(Sr^3{\p\Omega\over\p r}), \label{ang0}
\eeq
where $\Omega=v_\phi/r$, and 
\beq
S=\int_{-\infty}^\infty \rho\nu {\mathrm d}z\approx\Sigma\nu. \label{s}
\eeq
The second approximate equality in~(\ref{s}) holds if $\nu$ can be considered
independent of $z$.  The right hand side of (\ref{ang0}) is the divergence of the 
viscous angular momentum flux, and is derived most easily with a physical 
argument, as described in, e.g.\ Pringle (1981) or Frank et al.\ (2002)\footnote{If
you prefer a more formal derivation, the fastest way is to consult Landau and
Lifshitz (1959) chapter 15 (hereafter LL). Noting that the divergence of the
flow vanishes for a thin axisymmetric disk, the viscous stress $\sigma$
becomes (LL eq. 15.3)  
\bd 
\sigma_{ik}=\eta\left({\p v_i\over\p x_k}+{\p v_k\over\p x_i} \right), 
\ed 
where $\eta=\rho\nu$. This can be written in cylindrical or spherical coordinates 
using LL eqs. (15.15-15.18). The viscous force is 
\bd 
F_i={\p\sigma_{ik}\over\p x_k}={1\over\eta}{\p\eta\over\p x_k}\sigma_{ik}
+\eta\nabla^2 v_i, 
\ed
Writing the Laplacian in cylindrical coordinates, the viscous torque is then
computed  from the $\phi$-component of the viscous force by multiplying by $r$,
and is then integrated over $z$.}. 

Assume now that $\nu$ can be taken constant with height. For an isothermal disk
($T$ independent of $z$), this is equivalent to taking the viscosity parameter 
$\alpha$ independent of $z$. As long as we are not sure what causes the 
viscosity this is a reasonable simplification. Note, however, that recent
numerical simulations of magnetic turbulence suggest that the effective 
$\alpha$, and the rate of viscous dissipation per unit mass, are higher near the 
disk surface than near the midplane (c.f. section 
\ref{source}). While eq.\ (\ref{ang0}) is still valid for rotation rates $\Omega$
deviating from Keplerian (only the integration over disk thickness must be
justifiable), we now use the fact that $\Omega\sim r^{-3/2}$. Then
eqs.~(\ref{co}-\ref{ang0}) can be combined into a single equation for $\Sigma$:
\beq  r{\p\Sigma\over\p t}=3{\p\over\p r}[r^{1/2}{\p\over\p r}(\nu
\Sigma r^{1/2})] .\label{dif} 
\eeq
Under the same assumptions, eq. (\ref{vphi}) yields the mass flux $\dot M$ at any
point in the disk:
\beq \dot M=-2\pi r\Sigma v_r=6\pi r^{1/2}{\p\over\p r}(\nu\Sigma
r^{1/2}).\label{dm}\eeq 

Eq.\ (\ref{dif}) is the standard form of the {\it thin disk diffusion equation}. An
important conclusion from this equation is: in the thin disk limit, all the
physics which influences the time dependent behavior of the disk enters 
through one quantity only: the viscosity $\nu$. This is the main attraction
of the thin disk approximation.

\subsection{Steady thin disks}
In a steady disk ($\p/\p t=0$) the mass flux $\dot M$ is constant through the
disk and equal to the accretion rate onto the central object. From (\ref{dm})
we get the surface density distribution:
\beq 
\nu\Sigma={1\over 3\pi}\dot M\left[1-\beta\left({r_i\over r}\right)^{1/2}
\right], \label{ns}
\eeq
where $r_i$ is the inner radius of the disk and $\beta$ is a parameter
appearing through the integration constant. It is related to the flux of
angular momentum $F_J$ through the disk: 
\beq F_J=-\dot M\beta\Omega_ir_i^2 \label{fj}, \eeq 
where $\Omega_i$ is the Kepler rotation rate at the inner edge of the disk.
If the disk accretes onto an object with a rotation rate $\Omega_*$ {\it less}
than $\Omega_i$, the rotation rate $\Omega(r)$ as a function of distance $r$ 
jumps from $\Omega_*$ close to the star to the Kepler rate $\Omega_{\rm K}(r)$ 
in the disk. It then has a maximum at some distance close to the star (distance of the 
order of the disk thickness $H$). At this point the viscous stress (proportional to 
$\partial_r\Omega$) vanishes and the angular momentum flux is just the amount 
carried by the accretion flow. In (\ref{fj}) this corresponds to $\beta=1$, {\em 
independent} of $\Omega_*$ (Shakura and Sunyaev, 1973, Lynden-Bell and
Pringle, 1974). This is referred to as the standard or `accreting case'. The angular
momentum flux (equal to the torque on the accreting star), is then inward, 
causing the rotation rate of the star to increase (spinup).

\begin{figure}
\hfil \includegraphics[width=5.5 cm]{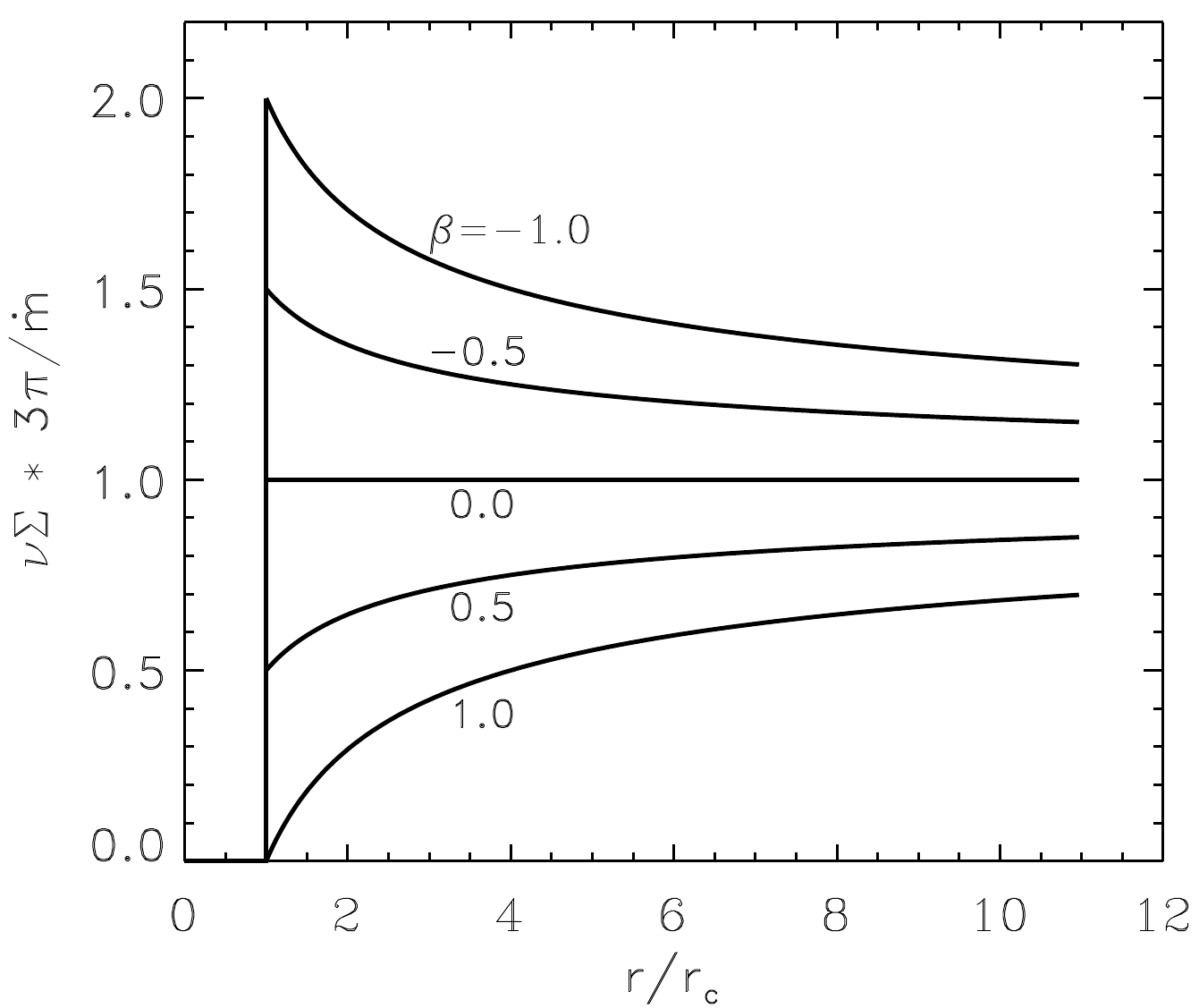}\hfil \includegraphics [width=5.5 cm]{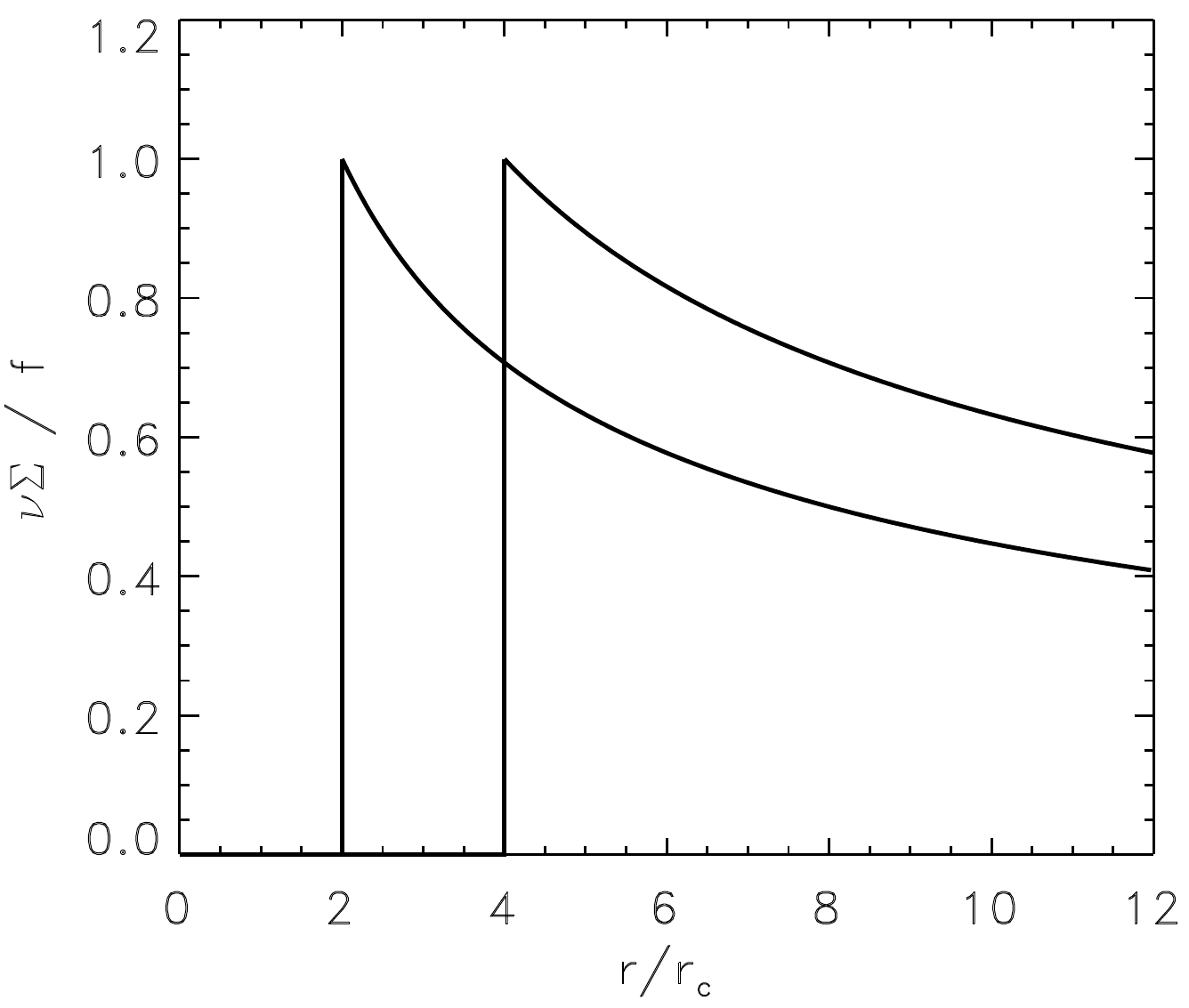}\hfil
\caption[]{Surface density $\nu\Sigma$ of a thin disc as a function of
  distance from the corotation radius $r_{\rm c}$, for a steady, thin
  viscous disc. Left: steady accretion at a fixed accretion rate $\dot
  m$, for inner edge of the disc at corotation. $\beta$ measures the
  angular momentum flux, $\beta=-1$ corresponding to the standard case
  of accretion on to a slowly rotation object. For $\beta < 0$ the
  angular momentum flux is outward (spindown of the star). Right:
  `quiescent disc' solutions with $\dot{m} = 0$ and a steady outward
  angular momentum flux due to a torque $f$ applied at the inner
  edge.\label{sigma}}
\end{figure}

For stars rotating near their maximum rate ($\Omega_*\approx\Omega_i$) 
and for accretion onto magnetospheres, which can
rotate faster than the disk, the situation is different (Sunyaev and Shakura 1977,
Popham and Narayan 1991, Paczy\'nski 1991, Bisnovatyi-Kogan 1993, 
Rappaport et al. 2004). If the inner edge of the disk is at the {\em corotation 
radius} $r_{\rm co}$, defined by
\beq \Omega_{\rm K}(r_{\rm co})=\Omega_*,\eeq
the viscous stress cannot be assumed to vanish there, and the thin disk 
approximation does not give a unique answer for the angular momentum 
flux parameter $\beta$. Its value is then determined by details of the 
hydrodynamics at $r_{\rm co}$, which need to be investigated separately.
Depending on the outcome of this investigation, values varying from 1 
(spinup, standard accreting case) to negative values (spindown) are possible. 
As (\ref{fj}) shows, the surface density at the inner edge of the disk depends
sensitively on the value of $\beta$ (see also Fig.\ \ref{sigma}). This plays a
role in the cyclic accretion process discussed in the next subsection.

\subsection{Magnetospheres, `propellering'  and `dead disks'}
Stars with magnetospheres, instead of spinning up by accretion of angular 
momentum from the disk, can actually {\it spin down} by interaction with the
disk, even while accretion is still going on. The thin disk approximation 
above covers this case as well (cf. figure \ref{sigma}). The surface 
density distribution is then of the form (\ref{ns}), but with $\beta<0$ (see also 
Spruit and Taam 1993, Rappaport et al. 2004). The angular momentum flux 
through the disk is outward, and the accreting star spins down. This is 
possible even when the interaction between the disk and the magnetosphere 
takes place {\it only} at the inner edge of the disk. Magnetic torques due interaction 
between the disk and the magnetosphere may exist at larger distances in the disk 
as well, but are not necessary for creating an outward angular momentum flux. 
Numerical simulations of disk-magnetosphere interaction (Miller and Stone 1997,
for recent work see Long et al. 2008 and references therein) give an interesting 
view of how such interaction may take place, presenting a picture that is very 
differently from what is assumed in the previous `standard' models. Among other
things, they show the interaction region to be quite narrow.

\begin{figure}
\hfil\includegraphics[width=12 cm, clip]{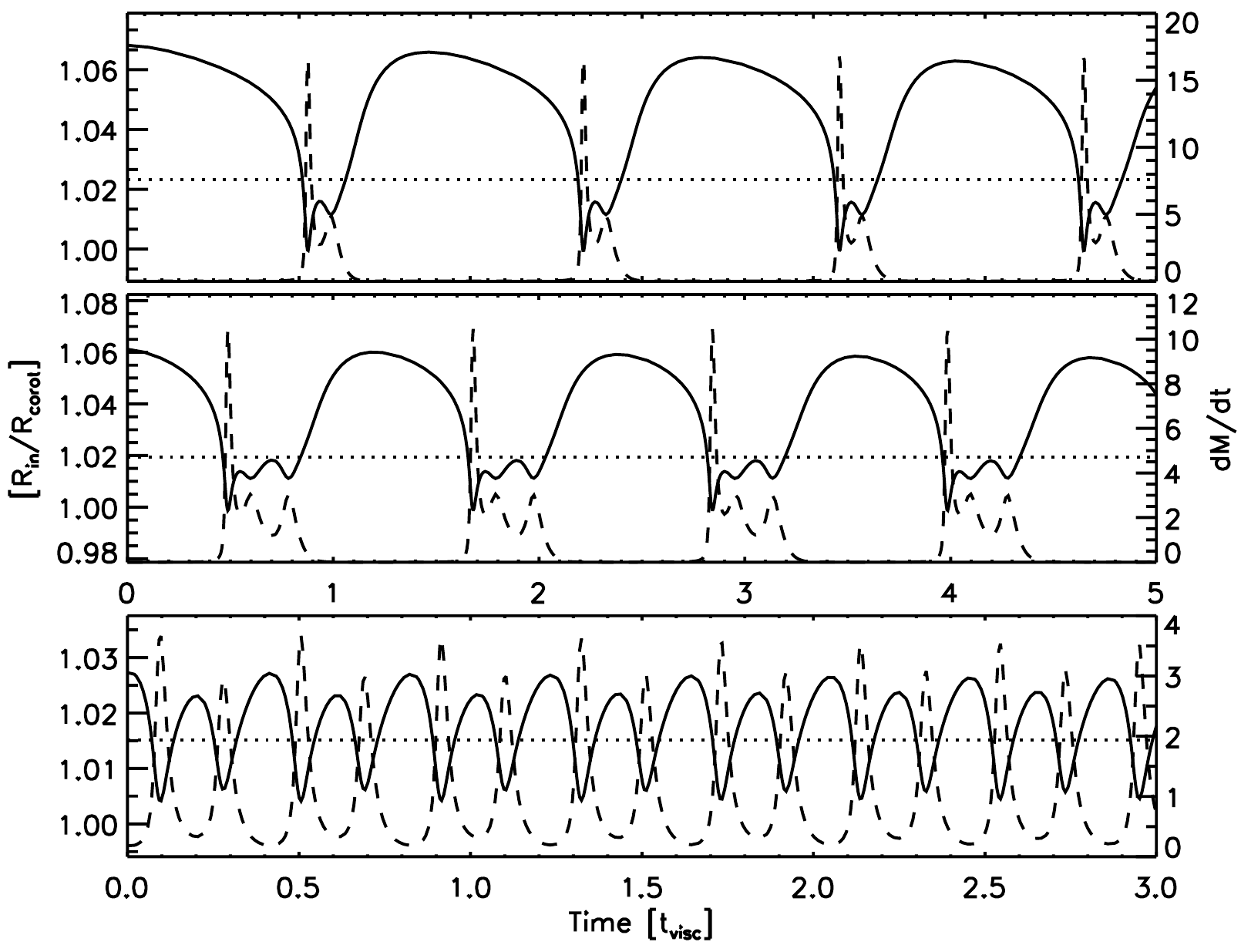}\hfil
\caption{Cyclic accretion due to interaction between a magnetosphere and a 
disk with inner edge near corotation. Solid: inner edge radius of the disk in 
units of the corotation radius; dashed: accretion rate (from D'Angelo and Spruit 
2010).}\label{cycles}
\end{figure}

Rapidly rotating magnetospheres play a role in some CVs and X-ray 
binaries, and probably also in protostellar accretion disks.  When 
the rotation velocity of the magnetosphere is larger than the Kepler velocity 
at the inner edge of the disk (i.e. when $r_{\rm i}> r_{\rm co}$),  the literature 
often assumes that the mass transfered from the secondary must be 
`flung out' of the binary system. This idea is idea called `propellering' 
(Illarionov and Sunyaev 1975). The term 
is then used as synonymous with the condition $r_{\rm i}> r_{\rm co}$. 
While a process like this is likely to happen when the rotation rate of the star 
is sufficiently large (the CV AE Aqr being an example) it is not necessary. It
is not possible either, when the difference in rotation velocity is too small to put the 
accreting mass on an escape orbit. This has been realized early on (Sunyaev 
and Shakura 1977). Instead of being flung out, accretion is halted and mass 
accumulates in the disk. The buildup of mass in the disk then leads to a following 
episode of accretion. In this phase the magnetosphere is compressed such 
that $r_{\rm i}< r_{\rm co}$. Instead 
of propellering, the result is {\em cyclic} accretion. For quantitative models 
see D'Angelo and Spruit (2010), an example is shown in Figure \ref{cycles}. The 
phase with $r_{\rm i}> r_{\rm co}$, during which accretion is halted is called 
`dead disk' phase in Sunyaev and Shakura (1977).

\subsection{Disk Temperature}
In this section I assume accretion onto not-too-rapidly rotating objects, so
that $\beta=1$ (eq. \ref{ns}). The surface temperature of the disk, which 
determines how much energy it loses by radiation, is governed by the 
energy dissipation rate in the disk, which in turn is given by the accretion 
rate. From the first law of thermodynamics we have
\beq \rho T{{\mathrm d}S\over {\mathrm d}t}=-{\mathrm div}{\mathbf F}+Q_{\rm v},
\eeq
where $S$ is the entropy per unit mass, $\mathbf F$ the heat flux (including
radiation and convection), and $Q_{\rm v}$ the 
viscous dissipation rate. For thin disks, where the advection of internal energy 
(or entropy) can be neglected, and for changes which happen on time scales 
longer than the
dynamical time $\Omega^{-1}$, the left hand side is small compared with the
terms on the right hand side. Integrating over $z$, the divergence turns
into a surface term and we get 
\beq 
2\sigma_{\rm r}T_{\rm s}^4=\int_{-\infty}^\infty Q_{\rm v}{\mathrm d}z,\label{bal}
\eeq 
where $T_{\rm s}$ is the surface temperature of the disk, $\sigma_{\rm r}$ is
Stefan-Boltzmann's radiation constant $\sigma_{\rm r}=a_{\rm r}c/4$, and the
factor 2 comes about because the disk has 2  radiating surfaces (assumed to
radiate like black bodies).
Thus the energy balance is {\it local} (for such slow changes): what is
generated by viscous dissipation inside the disk at any radius $r$ is also
radiated away from the surface at that position. The viscous dissipation rate
is equal to $Q_{\rm v}=\sigma_{ij}\p v_i/\p x_j$, where $\sigma_{ij}$
is the viscous stress tensor (see footnote in section 5), and this works
out\footnote{using, e.g. LL eq. 16.3}
to be 
\beq Q_{\rm v}=9/4\,\Omega^2\nu\rho.\label{qv}\eeq 
Eq.~(\ref{bal}), using (\ref{ns}) then  gives the surface temperature in terms of
the accretion rate: 
\beq 
\sigma_{\rm r} T_{\rm s}^4={9\over 8}\Omega^2\nu\Sigma={GM\over r^3}{3\dot
M\over 8\pi} \left[1-\left({r_i\over r}\right)^{1/2}\right]. \label{st}
\eeq
This shows that the surface temperature of the disk, at a given distance $r$
from a steady accreter, depends {\it only} on the product $M\dot M$, and not on
the highly uncertain value of the viscosity. For $r\gg r_i$ we have
\beq T_{\rm s}\sim r^{-3/4}. \eeq 

These considerations only tells us about the surface temperature. The
internal temperature in the disk is quite different, and depends on the
mechanism transporting energy to the surface. Because it is the internal
temperature that determines the disk thickness $H$ (and probably also the
viscosity), this transport needs to be considered in some detail for
realistic disk models. This involves a calculation of the vertical structure of
the disk. Because of the local (in $r$) nature of the balance between
dissipation and energy loss, such calculations can be done as a grid of models
in $r$, without having to deal with exchange of energy between neighboring
models. Schemes borrowed from stellar structure computations are used (e.g.
Meyer and Meyer-Hofmeister 1982, Cannizzo et al. 1988). 

An approximation to the temperature in the disk can be found  when a number
of additional assumptions is made.  As in stellar interiors, the energy transport
is radiative rather than convective at high temperatures. Assuming local
thermodynamic equilibrium (LTE, e.g. Rybicki and Lightman 1979), the temperature
structure of a radiative atmosphere is given, in the so-called Eddington 
approximation of radiative transfer (not to be confused with the Eddington limit) by:
\beq {\rmd\over\rmd\tau} \sigma_{\rm r} T^4={3\over 4}F.\label{edd}\eeq
The boundary condition that there is no incident flux from outside the atmosphere
yields the approximate condition
\beq \sigma_{\rm r} T^4(\tau=2/3)=F, \label{sb}\eeq
where $\tau=\int_z^\infty\kappa\rho \rmd z$ is the optical depth at geometrical
depth $z$, and $F$ the energy flux through the atmosphere. Assuming that most
of the heat is generated near the midplane (which is the case if $\nu$ is
constant with height), $F$ is approximately constant with height and equal to
$\sigma_{\rm r}T^4_{\rm s}$, given by (\ref{st}). Eq (\ref{edd}) then yields
\beq \sigma_{\rm r} T^4={3\over 4}(\tau+{2\over 3})F.\label{t4}\eeq
Approximating the opacity $\kappa$ as constant with $z$, the optical depth at the
midplane is $\tau=\kappa\Sigma/2$. If $\tau\gg1$, the temperature at the midplane
is then: 
\beq 
T^4={27\over 64}\sigma_{\rm r}^{-1}\Omega^2\nu\Sigma^2\kappa .\label{ts}
\eeq 
With the equation of state (\ref{es}), valid when radiation pressure is
small, we find for the disk thickness, using (\ref{ns}):  
\begin{eqnarray} 
{H\over r}=&({\cal R}/\mu)^{2/5}\left({\sigma_{\rm r}64/3\pi^2}
\right)^{-1/10}(\kappa/\alpha)^{1/10}(GM)^{-7/20}r^{1/20}(f\dot M)^{1/5} \cr
 =&5\,10^{-3}\alpha^{-1/10} r_6^{1/20} \left({M/M_\odot}\right)^{-7/20}
(f\dot M_{16})^{1/5}, \qquad (P_r\ll P)\label{hr} 
\end{eqnarray}
where $r_6=r/(10^6$ cm), $\dot M_{16}=\dot M/(10^{16}$g/s), and
\bd f=1-\left({r_i/ r}\right)^{1/2}. \ed
From this we conclude: i) that the disk is thin in X-ray binaries, $H/r<0.01$, ii)
the disk thickness is relatively insensitive to the parameters,
especially $\alpha$, $\kappa$ and $r$. It must be stressed, however, that this
depends fairly strongly on the assumption that the energy is dissipated in the
disk interior. If the dissipation takes place close to the surface, such as in
some magnetic reconnection models, the internal disk temperature will be much 
closer to the surface temperature  (Haardt et al. 1994, Di Matteo et al. 1999a and
references therein). The midplane temperature and $H$ are even smaller in such
disks than calculated from (\ref{hr}).

\subsection{A subtlety: viscous vs. gravitational energy release}
\label{visgrav}
The viscous dissipation rate per unit area of the disk, $W_{\rm
v}=(9/4)\Omega^2\nu\Sigma$ [cf. eq. \ref{st})] can be compared with the local rate
$W_{\rm G}$ at which gravitational energy is liberated in the accretion flow.
Since half the gravitational energy stays in the flow as orbital motion, we have
\beq W_{\rm G}= {1\over 2\pi r}{GM\dot M\over 2 r^2},\eeq
so that, in a steady thin disk accreting on a slowly rotating object (using eq. \ref{ns}):
\beq W_{\rm v}/W_{\rm G}=3f=3[1-(r_{\rm i}/r)^{1/2}]. \eeq
At large distances from the inner edge, the dissipation rate is {\em 3 times
larger than the rate of gravitational energy release}. This may seem odd, but
becomes understandable when it is realized that there is a significant flux of
energy through the disk associated with the viscous stress\footnote{See Landau 
\& Lifshitz section 16}. Integrating the viscous energy dissipation over the whole disk, 
one finds \beq \int_{r_i}^\infty 2\pi r W_{\rm v} \rmd r={GM\dot M\over 2 r_{\rm i}},\eeq 
as expected. That is, globally but not locally, half of the gravitational energy
is radiated from the disk while the other half remains in the orbital kinetic
energy of the material  accreting on the star. 

What happens to this remaining orbital energy depends on the nature of the
accreting object. If the object is a slowly rotating black hole, the orbital
energy is just swallowed by the hole. If it has a solid surface, the orbiting gas
slows down until it corotates with the surface, dissipating the orbital energy
into heat in a boundary layer. Unless the surface rotates close to the orbital
rate (`breakup'), the energy released in this way is of the same order as the
total energy released in the accretion disk. The properties of this boundary layer
are therefore critical for accretion onto neutron stars and white dwarfs. See also
section \ref{bl} and Inogamov and Sunyaev (1999).

\subsection{Radiation pressure dominated disks}
In the inner regions of disks in XRB, the radiation pressure can dominate over
the gas pressure, which results in a different expression for the disk 
thickness. The total pressure $P$ is 
\beq P=P_r+P_g={1\over 3}aT^4+P_g.\eeq
Defining a `total sound speed' by $c_{\rm t}^2=P/\rho$ the relation between 
temperature and disk thickness (\ref{h})  still holds, so $c_{\rm t}=\Omega H$. 
For $P_r\gg P_g$ we get from (\ref{ts}), with (\ref{st}) and $\tau\gg 1$:
\bd c H={3\over 8\pi}\kappa f\dot M, \ed
(where the rather approximate relation $\Sigma=2H\rho_0$ has been used). Thus,
\beq 
{H\over R}\approx{3\over 8\pi}{\kappa\over cR}f\dot M ={3\over 2}f{\dot M\over\dot
M_{\rm E}} , \label{he} 
\eeq
where $R$ is the stellar radius and $\dot M_{\rm E}$ the Eddington rate for this
radius. It follows that the disk becomes thick near the star, if the accretion
rate is near Eddington (though this is mitigated somewhat by the factor $f$). 
Accretion near the Eddington limit is evidently not geometrically
thin any more. In addition, other processes such as angular momentum loss from 
the disk by `photon drag' in the dense radiation field of the accreting star have to be 
taken into account. 

\section{Time scales in thin disks}
\label{times}
A number of time scales play a role in the behavior of disks. In a thin disk they 
differ by powers of the large factor $r/H$. The shortest of these is the dynamical 
time scale $t_{\rm d}$:
\beq t_{\rm d}=\Omega{\rm K}^{-1}=({GM/ r^3})^{-1/2}. \eeq
The time scale for radial drift through the disk over a distance of order $r$ is
the viscous time scale:
\beq 
t_{\rm v}=r/(-v_r)={2\over 3}{rf\over\nu}={2f\over 3\alpha\Omega}({r\over H})^2,
\eeq
(using (\ref{dm} and (\ref{ns}), valid for steady accretion). This is of the order 
$(r/H)^2$ longer than the dynamical time scale, except near the inner edge 
of the disk, where $f\downarrow 0$, andF this time scale formally drops to zero as a
consequence of the thin disk approximations. If more physics were included, the
viscous time scale would stay above the dynamical time scale.

Intermediate are
{\em thermal} time scales. If $E_{\rm t}$ is the thermal energy content (enthalpy)
of the disk per unit of surface area, and $W_{\rm v}= (9/4)\Omega^2 \nu\Sigma$ the
heating rate by viscous dissipation, we can define a heating time scale:
\beq t_{\rm h}=E_{\rm t}/W_{\rm v}. \eeq
In the same way, a cooling time scale is defined by the energy content and the
radiative loss rate:
\beq t_{\rm c}=E_{\rm t}/(2\sigma_{\rm r} T_{\rm s}^4).\eeq
For a thin disk, the two are equal since the viscous energy dissipation is locally
balanced by radiation from the two disk surfaces. [In thick disks (ADAFs), this
balance does not hold, since the advection of heat with the 
accretion flow is not negligible. In this case $t_{\rm c}> t_{\rm h}$ (see section
\ref{hydro})]. Thus, we can replace both time scales by a single thermal time scale
$t_{\rm t}$, and find, with (\ref{qv}):
\beq 
t_{\rm t}={1\over W_{\rm v}}\int_{-\infty}^\infty{\gamma P\over\gamma -1}\rmd z,
\eeq
where the enthalpy of an ideal gas of constant ratio of specific heats $\gamma$
has been used. Leaving out numerical factors of order unity, this yields
\beq t_{\rm t}\approx{1\over\alpha\Omega}.\eeq
That is, the thermal time scale of the disk is independent of most of the disk
properties and of the order $1/\alpha$ times longer than the dynamical time scale.
This independence is a consequence of the $\alpha$-parameterization used. If
$\alpha$ is not a constant, but dependent on disk temperature for example, the
dependence of the thermal time scale on disk properties will become apparent
again.

If, as seems likely from observations, $\alpha$ is generally $<1$, we have in thin
disks the ordering of time scales:
\beq t_{\rm v}\gg t_{\rm t}>t_{\rm d}.\eeq

\section{Comparison with CV observations}
The number of meaningful quantitative tests between the theory of disks and
observations is somewhat limited since in the absence of a theory for $\nu$,
it is a bit meagre on predictive power. The most detailed information perhaps
comes from modeling of CV outbursts. 

Two simple tests are possible (nearly) independently of $\nu$. These are the
prediction that the disk is geometrically quite thin (eq.~\ref{hr}) and
the prediction that the surface temperature $T_{\rm s}\sim r^{-3/4}$ in a steady
disk. The latter can be tested in a subclass of the CV's that do not show
outbursts, the nova-like systems, which are believed to be approximately
steady accreters. If such a system is also eclipsing, eclipse mapping techniques
can be used to derive the brightness distribution with $r$ in the disk (Horne,
1985, 1993). If this is done in a number of colors so that bolometric
corrections can be made, the results (e.g. Rutten et al. 1992) show in
general a {\it fair} agreement with the $r^{-3/4}$ prediction. Two deviations
occur: i) a few systems show significantly flatter distributions than
predicted, and ii) most systems show a `hump' near the outer edge of the
disk. The latter deviation is easily explained, since we have not taken into
account that the impact of the stream additionally heats the outer edge of the disk.
Though not important for the total light from the disk, it is an important
local contribution near the edge. 

Eclipse mapping of Dwarf Novae in quiescence gives a quite different picture.
Here, the inferred surface temperature profile is often nearly flat (e.g. Wood et
al. 1989a, 1992). This is understandable however since in quiescence the mass flux
depends strongly on $r$. In the inner parts of the
disk it is small, near the outer edge it is close to its average value. With
eq.~(\ref{st}), this yields a flatter $T_{\rm s}(r)$. The lack of light from the
inner disk is compensated during the outburst, when the accretion rate in the
inner disk is higher than average (see Mineshige and Wood 1989 for a more
detailed comparison). The effect is also seen  in the 2-dimensional
hydrodynamic simulations of accretion in a binary by R\'o\.zyczka and Spruit
(1993). These simulations show an outburst during which the accretion
in the inner disk is enhanced, between two episodes in which mass accumulates
in the outer disk.

\subsection{Comparison with LMXB observations: irradiated disks}
In low mass X-ray binaries a complication arises because of the much higher
luminosity of the accreting object. Since a neutron star is roughly 1000 times
smaller than a white dwarf, it produces 1000 times more luminosity for a given
accretion rate.

Irradiation of the disk by the central source (the accreting star plus inner disk) 
leads to a different surface temperature than predicted by (\ref{st}). The central 
source radiates nearly the total accretion luminosity $GM\dot M/R$ (assuming
sub-Eddington accretion, see section 2). If the disk is {\it concave}, it will
intercept some of this luminosity. If the central source is approximated as a
point source the irradiating flux on the disk surface is
\beq F_{\rm irr}={ \epsilon\over 2} {GM\dot M\over 4\pi Rr^2}, \eeq
where $\epsilon$ is the angle between the disk surface and the direction from a
point on the disk surface to the central source:
\beq \epsilon={\mathrm d}H/{\mathrm d}r-H/r.\label{fir}\eeq
The disk is concave if $\epsilon$ is positive. We have
\bd 
{F_{\rm irr}\over F}={1\over 3} {\epsilon\over f} {r\over R},
\ed
where $F$ is the flux generated internally in the disk, given by
(\ref{st}). On average, the angle $\epsilon$ is of the order of the aspect
ratio $\delta=H/r$. With $f\approx 1$, and our fiducial value $\delta\approx
5\,10^{-3}$, we find that irradiation in LMXB dominates for $r \gapprox 10^{9}$cm. This
is compatible with observations (for reviews see van Paradijs and McClintock
1993), which show that the optical and UV are dominated by reprocessed 
radiation from the innermost regions. 

When irradiation by an external source is included
in the thin disk model, the surface boundary condition of the radiative transfer
problem, equation (\ref{sb}) becomes
\beq 
\sigma_{\rm r} T_{\rm s}^4=F + (1-a)F_{\rm irr}, \label{sti}
\eeq
where $a$ is the X-ray albedo of the surface, that is, $1-a$ is the fraction of the
incident flux that is absorbed in the {\it optically thick} layers of the disk
(photons absorbed higher up only serve to heat up the corona of the disk)\footnote{Incorrect derivations exist in the literature, in which the effect 
of irradiation is treated like an energy flux added to the internal viscous 
dissipation inside the disk rather than incident on the surface.}. The
surface temperature $T_{\rm s}$ increases in order to compensate for the
additional incident heat flux. The magnitude of the incident flux is sensitive 
to the assumed disk shape $H(r)$, as well as on the assumed shape (plane or 
spherical, for example) of the central X-ray emitting region. 

The disk thickness 
depends on temperature, and thereby also on the irradiation. It turns out, 
however, that this dependence on the irradiating flux is small, if the disk is 
optically thick and the energy transport is by radiation (Lyutyi and Sunyaev 
1976). To see this, integrate (\ref{edd}) with the modified 
boundary condition (\ref{sti}). This yields
\beq
\sigma_{\rm r} T^4={3\over 4}F(\tau+{2\over 3})+(1-a)F_{\rm irr}.
\eeq
Thus the irradiation adds an additive constant to $T^4(z)$. At the midplane, this
constant has much less effect than at the surface. For the midplane temperature
and the disk thickness to be affected significantly, it is necessary that
\beq F_{\rm irr}/F \gapprox \tau.\eeq
The reason for this weak dependence of the midplane conditions on irradiation is
the same as in radiative envelopes of stars, which are also insensitive to the
surface boundary condition. The situation is different for convective disks.
As in fully convective stars, the adiabatic stratification then causes the
conditions at the midplane to depend much more directly on the surface temperature.
The outer parts of the disks in LMXB with wide orbits are in fact convective, hence their
thickness more directly affected by irradiation. This leads to the possibility of {\em 
shadowing}, with irradiated, vertically extended regions blocking irradiation of the 
disk outside. This plays an observable role in protostellar disks (Dullemond et al. 2001). 

\subsection{Observational evidence of disk thickness}
From the paucity of sources in which eclipses of the central source by the 
companion are observed one deduces that the companion is barely or not at all 
visible as seen from the inner disk. Apparently some parts of the disk
are much thicker than expected from the above arguments. This is consistent
with the observation that a characteristic modulation of the optical
light curve indicative of irradiation of the secondary's surface by the X-rays is
not very strong in LMXB (an exception being Her X-1, which has an atypically large
companion). The place of the eclipsing systems is taken by the so-called
`Accretion Disk Corona' (ADC) systems, where shallow eclipses of a rather
extended X-ray source are seen instead of the expected sharp eclipses of the inner
disk (for reviews of the observations, see Lewin et al. 1995). The conclusion is
that there is an extended X-ray scattering `corona' above the disk. It scatters a
few per cent of the X-ray luminosity.

What causes this corona and the large inferred thickness of the disk? The
thickness expected from disk theory is a rather stable
small number. To `suspend' matter at the inferred height of the disk forces are
needed that are much larger than the pressure forces available in an
optically thick disk. A thermally driven wind, produced by X-ray heating of the
disk surface, has been invoked (Begelman et al. 1983,  Schandl and
Meyer 1994, 1997). For other explanations, see van Paradijs and McClintock (1995).
Perhaps a magnetically driven wind from the disk (e.g. Bisnovatyi-Kogan \& 
Ruzmaikin 1976, Blandford \& Payne 1982), such as inferred for protostellar
objects (cf.\ Lee et al. 2000) can explain both the shielding of the
companion and the scattering. Such a model would resemble magnetically driven
wind models for the broad-line region in AGN (e.g. Emmering et al., 1992, K\"onigl
and Kartje 1994). A promising possibility is that the reprocessing region consists of 
matter `kicked up' at the disk edge by the impact of the mass transferring
stream (Meyer-Hofmeister et al. 1997, Armitage and Livio 1998, Spruit et al.
1998). This produces qualitatively the right dependence of X-ray absorption on
orbital phase in ADC sources, and the light curves of the so-called supersoft
sources.

\subsection{Transients}
Soft X-ray transients (also called X-ray Novae) are believed to be binaries
similar to the other LMXB, but somehow the accretion is episodic, with very
large outbursts recurring on time scales of decades (sometimes years). Most of
these transients turn out to be black hole candidates (see Lewin et al. 1995 for a
review). As with the Dwarf Novae, the time dependence of
the accretion can in principle be exploited in transients  to derive
information on the disk viscosity, assuming that the outburst is caused by an
instability in the disk. The closest relatives of soft transients among the
White Dwarf plus main sequence star systems are probably the WZ Sge stars
(van Paradijs and Verbunt 1984, Kuulkers et al. 1996), which show 
(in the optical) similar outbursts with similar recurrence times (cf. Warner 
1987, O'Donoghue et al. 1991). Like the soft transients, they have low mass 
ratios ($q<0.1$). For a given angular momentum loss, systems with low mass 
ratios have low mass transfer rates, so the speculation is that the peculiar 
behavior of these systems is somehow connected in both cases with a low mean 
accretion rate. 

\subsubsection{Transients in quiescence}
X-ray transients in quiescence (i.e. after an outburst) usually show a very low 
X-ray luminosity. The mass transfer rate from the secondary in quiescence can be
inferred from  the optical emission. This shows the characteristic `hot spot',
known from other systems to be the location where the mass transferring stream
impacts on the edge of an accretion disk (e.g. van Paradijs and McClintock 1995).
These observations thus show that a disk is present in quiescence, while the mass
transfer rate can be measured from the brightness of the hot spot. If this disk
were to extend to the neutron star with constant mass flux, the predicted X-ray
luminosity would be much higher than observed. This has traditionally been
interpreted as a consequence of the fact that in transient systems, the accretion
is not steady. Mass is stored in the outer parts and released  by a disk
instability (e.g. King 1995, Meyer-Hofmeister and Meyer 1999) producing the X-ray
outburst. During quiescence, the accretion rate onto the compact object is much
smaller than the mass transfer from the secondary to the disk.

\subsection{Disk Instability}
\label{inst}
The most developed model for outbursts is the disk instability model of Osaki
(1974), H\= oshi (1979), Smak (1971, 1984), Meyer and Meyer-Hofmeister 
(1981), see also King (1995), Osaki (1993). In
this model the instability that gives rise to cyclic accretion is due to a
temperature dependence of the viscous stress. In any local process that causes
an effective viscosity, the resulting $\alpha$- parameter will be a function of
the main dimensionless parameter of the disk, the aspect ratio $H/r$. If this
is a sufficiently rapidly increasing function, such that $\alpha$ is large in
hot disks and low in cool disks, an instability results by the following
mechanism. Suppose we start the disk in a stationary state at the mean
accretion rate. If this state is perturbed by a small temperature increase,
$\alpha$ goes up, and by the increased viscous stress the mass flux $\dot M$
increases. By (\ref{st}) this increases the disk temperature further,
resulting in a runaway to a hot state. Since $\dot M$ is larger now than the
average, the disk empties partly, reducing the surface density and the central
temperature (eq. \ref{ts}). A cooling front then transforms the disk to a cool
state with an accretion rate below the mean. The disk in this model switches
back and forth between hot and cool states. By adjusting the value of $\alpha$ in 
the hot and cool states, or by adjusting the functional dependence of $\alpha$ on
$H/r$, outbursts are obtained that agree reasonably with the observations of
soft transients (Lin and Taam 1984, Mineshige and Wheeler, 1989). A rather
strong increase of $\alpha$ with $H/r$ is needed to get the observed long
recurrence times. 

Another possible mechanism for instability has been found in 2-D numerical
simulations of accretion disks (Blaes and Hawley 1988, R\'o\.zyczka and
Spruit 1993). The outer edge of a disk is found, in these simulations, to
become dynamically unstable to an oscillation which grows into a strong
eccentric perturbation (a crescent shaped density enhancement which rotates at
the local orbital period). Shock waves generated by this perturbation spread
mass over most of the Roche lobe; at the same time the accretion rate onto the
central object is strongly enhanced. This process is different from the
Smak-Osaki-H\=oshi mechanism, since it requires 2 dimensions, and does not
depend on the viscosity (instead, the internal dynamics in this instability
{\it generates} the effective viscosity that causes a burst of accretion).

\subsection{Other Instabilities}
Instability to heating/cooling of the disk can be the due to several effects.
The cooling rate of the disk, if it depends on temperature in an appropriate
way, can cause a thermal instability like that in the interstellar medium.
Other instabilities may result from the dependence of viscosity on conditions
in the disk. For a general treatment see Piran (1978), for a shorter
discussion see Treves et al. (1988).

\section{Sources of Viscosity}
\label{source}
The high Reynolds number of the flow in accretion disks (of the order
$10^{14}$ in the outer parts of a CV disk) would, to most fluid dynamicists,
seem an amply sufficient condition for the occurrence of hydrodynamic
turbulence (see also the discussion in section \ref{turbu}). A theoretical 
argument against such turbulence often used in
astrophysics (Kippenhahn and Thomas 1981, Pringle 1981) is that in cool disks
the gas moves almost on Kepler orbits, which are quite stable (except for the
orbits that get close to the companion or near a black hole). This stability is 
related to the known stabilizing effect that rotation has on hydrodynamical 
turbulence (Bradshaw 1969, Lesur \& Longaretti 2005). A (not very 
strong) observational argument is that hydrodynamical turbulence as 
described above would produce an $\alpha$ that does not depend on the 
nature of the disk, so that all objects should have the same value, which is 
not what observations show. From the modeling of CV outbursts one knows,
for example, that $\alpha$ probably increases with temperature (more
accurately, with $H/r$, see previous section). Also, there are indications
from the inferred life times and sizes of protostellar disks (Strom et al.
1993) that $\alpha$ may be rather small there, $\sim 10^{-3}$, whereas in
outbursts of CV's one infers values of the order $0.1-1$. 
 
Among the processes that have been proposed repeatedly as sources of 
viscosity is convection due to a vertical entropy gradient (e.g. Kley et
al. 1993), which may have some limited effect in convective parts of disks.
Rotation rate varies somewhat with height in a disk, due to the radial temperature 
gradient (`baroclinicity'). Modest amounts of turbulence have been reported
from instability of this form of differential rotation (Klahr \& Bodenheimer 2003, 
Lesur and Papaloizou 2010). 

Another class are {\it waves} of various kinds. Their effect can be global,
that is, not reducible to a local viscous term because by traveling across
the disk they can communicate torques over large distances. For example, waves
set up at the outer edge of the disk by tidal forces can travel inward and by
dissipating there can effectively transport angular momentum {\it outward}
(e.g. Narayan et al. 1987,  Spruit et al. 1987). A nonlinear version of this
idea are selfsimilar spiral shocks, observed in numerical simulations (Sawada
et al. 1987) and studied analytically (Spruit 1987). Such shocks can
produce accretion at an effective $\alpha$ of $0.01$ in hot disks, but are
probably not very effective in disks as cool as those in CV's and XRB. 

A second non-local mechanism is provided by a magnetically accelerated {\it
wind} originating from the disk surface (Blandford 1976, Bisnovatyi-Kogan and
Ruzmaikin 1976, Lovelace 1976, for an introduction see Spruit 1996). In principle,
such winds can take care of {\it all} the angular momentum loss needed to make
accretion possible in the absence of a viscosity (Blandford 1976, K\"onigl 1989).
The attraction of this idea is that
magnetic winds are a strong contender for explaining the strong outflows and
jets seen in some protostellar objects and AGN. It is not yet clear however if,
even in these objects, the wind is actually the main source of angular
momentum loss. 

In sufficiently cool or massive disks, selfgravitating
instabilities of the disk matter can produce internal friction. Pacz\'ynski
(1978) proposed that the resulting heating would limit the instability
and keep the disk in a well defined moderately unstable state. The angular
momentum transport in such a disk has been studied numerically (e.g.
Gammie 1997, Ostriker et al. 1999). Disks in CVs 
and XRB are too hot for selfgravity to play a role, but it can be important
in protostellar disks (cf.\ Rafikov 2009).

\subsection{magnetic viscosity}
\label{magdisk}
Magnetic forces can be very effective at transporting angular momentum. If it
can be shown that the shear flow in the disk produces some kind of small
scale fast dynamo process, that is, some form of magnetic turbulence, an effective
$\alpha\sim O(1)$ would be expected (Shakura and Sunyaev 1973). Numerical 
simulations of initially weak magnetic fields in accretion disks show that this does 
indeed happen in sufficiently ionized disks (Hawley et al. 1995,  Brandenburg et al.
1995, Balbus 2003). These show a small scale magnetic field with azimuthal
component dominating (due to stretching by differential rotation). The angular 
momentum transport is due to
magnetic stresses, the fluid motions induced by the magnetic forces contribute
only little to the angular momentum transport. In a perfectly conducting plasma
this turbulence can develop from an arbitrarily small initial field through
magnetic shear instability (also called magnetorotational instability, Velikhov
1959, Chandrasekhar 1961, Balbus and Hawley 1991). The significance of 
this instability is that it shows that at large conductivity accretion disks must be
magnetic. 

The actual form of the highly time dependent small scale magnetic field
which develops can only be found from numerical simulations. The simplest case 
thought to be representative of the process considers a vertically unstratified disk 
(component of gravity perpendicular to the disk ignored), of which only a radial
extent of a few times $H$ is included (cf. the causality argument above). Different 
simulations have yielded somewhat different values for the effective viscosity, 
contrary to the expectation that at least in this simplest form the process should yield 
a unique value. 

The nature of these differences has been appreciated only recently. 
Fromang et al. (2007a) show that the results do not converge with increasing 
numerical resolution (the effective $\alpha$ found appears to decrease indefinitely). 
The result also depends on the magnetic Prandtl number $P_{\rm m}= 
\nu/\eta_{\rm m}$, the ratio of viscosity to magnetic diffusivity $\eta_{\rm m}$. 
No turbulence is found for  $P_{\rm m}<1$ Fromang et al. (2007b). The significance of 
these findings is not quite obvious yet, but they clearly contradict the commonly 
assumed `cascade' picture of MHD turbulence (taken over from what is assumed 
in hydrodynamic turbulence, where the large scale behavior of turbulence appears 
to converge with numerical resolution). 

Perhaps the behavior found in this case is 
related to the highly symmetric nature of the idealized problem.  This is suggested 
by the finding (Davis et al. 2010, Shi et al. 2010) that effective viscosity appears 
to converge again with increasing numerical resolution when vertical stratification 
is included. By inference, the process defining the state of magnetic turbulence is 
different in this case from that in the unstratified case. 
It can plausibly be attributed to magnetic buoyancy instabilities (Shi et al. 2010). 
This would make the process somewhat analogous to the mechanism operating 
the solar cycle (Spruit 2010). Magnetic turbulence also behaves different if the net 
vertical magnetic flux 
crossing the simulated box does not vanish. (This flux is a conserved quantity set 
by the initial conditions). Magnetic turbulence then develops more easily (at lower
numerical resolution), and appears to converge as the numerical resolution is
increased.

The consequences of these new findings for angular momentum transport in 
disks are still to be settled (c.f. the insightful discussion in Lesur and Ogilvie 
2008).

\subsection{viscosity in radiatively supported disks}
A disk in which the radiation pressure $P_{\rm r}$ dominates must be optically
thick (otherwise the radiation would escape).  The radiation pressure then adds to
the total pressure. The pressure is larger than it would be, for a given
temperature, if only the gas pressure were effective. If the viscosity is then
parametrized by (\ref{alf}), it turns out (Lightman and Eardley, 1974) that the
disk is locally unstable. An increase in temperature increases the radiation
pressure, which increases the viscous dissipation and the temperature, leading to
a runaway. This has raised the question whether the radiation pressure should be
included in the sound speed that enters expression (\ref{alf}). If it is left out,
a lower viscosity results, and there is no thermal-viscous runaway. Without
knowledge of the process causing the effective viscous stress, this question can
not be answered.  Sakimoto and Coroniti (1989) have shown, however, that if the
stress is due to some form of magnetic turbulence, it most likely scales with the
gas pressure alone, rather than the total pressure. Now that it seems likely, from
the numerical simulations, that the stress is indeed magnetic, there is reason to
believe that in the radiation pressure-dominated case the effective viscosity will
scale as $\nu\sim \alpha P_{\rm g}/(\rho\Omega)$, making such disks thermally 
stable, as indicated by the results of Hirose et al. (2009).

\section{Beyond thin disks}
Ultimately, much of the progress in developing useful models of accretion
disks will depend on detailed numerical simulations in 2 or 3 dimensions. In
the disks one is interested in, there is usually a large range in length
scales (in LMXB disks, from less than the $10$ km neutron star
radius to the more than $10^{5}$ km orbital scale). Correspondingly, there is
a large range in time scales that have to be followed. This not technically
possible at present and in the foreseeable future. In numerical simulations
one is therefore limited to studying in an approximate way aspects that are
either local or of limited dynamic range in $r,t$ (for examples, see Hawley
1991, Armitage 1998, De Villiers et al. 2003, Hirose et al. 2004). For this reason, 
approaches have been used that relax the strict thin disk framework somewhat without
resorting to full simulations.  Some of the physics of
thick disks can be included in a fairly consistent way in the  `slim disk'
approximation (Abramowicz et al., 1988). The so-called Advection Dominated
Accretion Flows (ADAFs) are related to this approach (for a review see Yi 1998). 
They are discussed in sections \ref{radv}, \ref{hydro}, \ref{isaf} below).

\subsection{Boundary layers}
\label{bl}
In order to accrete onto a star rotating at the rate $\Omega_*$, the disk matter
must dissipate an amount of energy given by
\beq 
{GM\dot M\over 2R}\left[1-\Omega_*/\Omega_k(R)\right]^2. \label{wb}
\eeq
The factor in brackets measures the kinetic energy of the matter at the inner
edge of the disk ($r=R$), in the frame of the stellar surface. Due to this
dissipation the disk inflates into a `belt' at the equator of the star, of
thickness $D$ and radial extent of the same order. Equating the radiation
emitted from the surface of this belt to (\ref{wb}) one gets for the surface
temperature $T_{\rm sb}$ of the belt, assuming optically thick conditions and a
slowly rotating star ($\Omega_*/\Omega_k\ll1$):
\beq 
{GM\dot M\over 8\pi R^2 D}=\sigma_{\rm r} T_{\rm sb}^4
\eeq
To find the temperature inside the belt and its thickness, use eq.~(\ref{t4}).
The value of the surface temperature is higher, by a factor of the order
$(R/D)^{1/4}$, than the simplest thin disk estimate (\ref{st}, ignoring the
factor $f$). In practice, this works out to a factor of a few.
The surface of the belt is therefore not very hot. The situation is quite
different if the boundary layer is not optically thick (Pringle and Savonije
1979). It then heats up to much higher temperatures. 

Analytical methods to
obtain the boundary layer structure have been used by Regev and Hougerat
(1988), numerical solutions of the slim disk type by Narayan and Popham (1993),
Popham (1997), 2-D numerical simulations by Kley (1991). These considerations are
primarily relevant for CV disks; in accreting neutron stars, the dominant effects
of radiation pressure have to be included. More analytic progress on the structure
of the boundary layer between a disk and a neutron star and the way in which it
spreads over the surface of the star has been reported by Inogamov and Sunyaev 
(1999).

\section{Radiative efficiency of accretion disks}
In a thin accretion disk, the time available for the accreting gas to radiate 
away the energy released by the viscous stress is the accretion time,
\beq t_{\rm acc}\approx{1\over\alpha\Omk}({r\over H})^2, \eeq
where $\alpha$ is the dimensionless viscosity parameter, $\Omk$ the local 
Keplerian rotation  rate, $r$ the distance from the central mass, and $H$ the 
disk thickness (see Frank et al. 2002 or section \ref{times}). For a 
thin disk, $H/r\ll 1$, this time is much longer than the thermal time scale 
$t_{\rm t}\approx 1/(\alpha\Om)$ (section \ref{times}). There is then enough 
time for a local balance 
to hold between viscous dissipation and radiative cooling. For the accretion 
rates implied in observed systems the disk is then rather cool, which then 
justifies the starting assumption $H/r\ll 1$. 

This argument is somewhat circular, of course, since the accretion time is long 
enough for effective cooling only if the disk is assumed to be thin to begin 
with. Other forms of accretion disks may exist, even at the same accretion 
rates, in which the cooling is ineffective compared with that of standard 
(geometrically thin, optically thick) disks. In the following sections, we 
consider such forms of accretion and the conditions under which are to be 
expected.

Since radiatively inefficient disks tend to be thick, $H/r\sim O(1)$, they are 
sometimes called `quasi-spherical'. However, this does {\em not} mean that a 
spherically symmetric accretion model would be a reasonable approximation. 
The crucial difference is that the flow has angular momentum. The inward flow 
speed is governed by the rate at which angular momentum can be transferred 
outwards, rather than by gravity and pressure gradient as in the Bondi accretion
problem mentioned above. With $H/r\sim{\cal O}(1)$, the accretion time 
scale, $t_{\rm acc}\sim 1/(\alpha\Om)$ is still longer than the accretion time scale 
in the spherical case, $1/\Omega$, (unless the viscosity parameter $\alpha$ is
as large as $O(1)$). The dominant velocity component is azimuthal rather than 
radial, and the density and optical depth are much larger than in the spherical 
case.

It turns out that there are two kinds of radiatively inefficient disks, the 
optically thin and optically thick varieties. A second distinction occurs 
because accretion flows are different for central objects with a solid surface 
(neutron stars, white dwarfs, main sequence stars, planets), and those without 
(i.e. black holes). Start with optically  thick flows. 

\section{Radiation supported radiatively inefficient accretion}
\label{radv}
If the energy loss by radiation is small, the gravitational energy release 
$W_{\rm grav}\approx GM/(2r)$ is converted into enthalpy of the gas and 
radiation field\footnote{This assumes that a fraction $\sim 0.5$ of the
gravitational potential energy stays in the flow as orbital kinetic energy. This
is only an approximation, see also section \ref{hydro}.}
\beq 
{1\over 2}{GM\over r}={1\over\rho}[{\gamma\over\gamma-1}P_{\rm g}+ 
4P_{\rm r}], \label{tv}
\eeq
where an ideal gas of constant ratio of specific heats $\gamma$ has been 
assumed, and $P_{\rm r}={1\over 3}aT^4$ is the radiation pressure. In terms of 
the virial temperature $T_{\rm vir}=GM/({\cal R}r)$, and assuming $\gamma=5/3$, 
appropriate for a fully ionized gas (see section \ref{tvir}), this can be written
as 
\beq {T\over T_{\rm vir}}=[5+8{P_{\rm r}\over P_{\rm g}}]^{-1}. \label {ttv1} \eeq
Thus, for radiation pressure dominated accretion, $P_{\rm r}\gg P_{\rm g}$, the 
temperature is much less than the virial temperature: much of the accretion 
energy goes into photon production instead of heating. By hydrostatic equilibrium
the disk thickness is given by (cf section \ref{height})
\beq  H\approx [(P_{\rm g}+P_{\rm r})/\rho]^{1/2}/\Om,\eeq
With (\ref{ttv1}) this yields
\beq H/r\sim O(1).\eeq
In the limit $P_{\rm r}\gg P_{\rm g}$, the flow is therefore geometrically thick.
Radiation pressure then supplies a non-negligible fraction of the
support of the gas in the radial direction against gravity (the remainder being 
provided by rotation). 

For $P_{\rm r}\gg P_{\rm g}$, (\ref{tv}) yields
\beq {GM\over 2r}={4\over 3}{aT^4\over\rho}.\label{vir}\eeq
The radiative energy flux, in the diffusion approximation, is (eq. \ref{edd})
\beq 
F={4\over 3}{\rmd\over \rmd\tau}\sigma_{\rm r} T^4\approx {4\over 3}
{\sigma_{\rm r} T^4\over \tau}.
\eeq
Hence
\beq  F={1\over 8}{GM\over rH}{c\over\kappa}=F_{\rm E}{r\over 8H},\label{flux}\eeq
where $F_{\rm E}=L_{\rm E}/(4\pi r^2)$ is the local Eddington flux. Since 
$H/r\approx 1$, a radiatively inefficient, radiation pressure dominated accretion
flow has a luminosity of the order of the Eddington luminosity. 

The temperature depends on the accretion rate and the viscosity $\nu$ assumed. 
The accretion rate is of the order $\dot M\sim 3\pi\nu\Sigma$ (cf. eq. (\ref{ns})), 
where $\Sigma=\int\rho\rmd z$ is the surface mass density. In units of the 
Eddington rate $\dot M_{\rm E}$, eq.\ (\ref{me}), we get
\beq \dot m\equiv\dot M/\dot M_{\rm E}\approx {\nu\rho\kappa/c},\label{dotsm} \eeq
where $H/r\approx 1$ has been used\footnote{The definition of 
$\dot M_{\rm E}$ differs by factors of order unity 
between different authors. It depends on the assumed efficiency $\eta$ of 
conversion of gravitational energy $GM/R$ into radiation. For accretion onto black
holes a more realistic value is of order $\eta=0.1$, for accretion onto neutron
stars $\eta\approx 0.4$, depending on the radius of the star.}.
Assume that the viscosity scales with the  gas pressure:
\beq \nu=\alpha{P_{\rm g}\over\rho\Omk}, \eeq
instead of the total pressure $P_{\rm r}+P_{\rm g}$. This is the form that is 
likely to hold if the angular momentum transport is due to a small-scale 
magnetic field (Sakimoto and Coroniti, 1989, Turner 2004). Then with (\ref{vir}),
(\ref{dotsm}) we have (up to a numerical factor of ${\cal O}(1))$
\beq 
T^5\approx {(GM)^{3/2}\over r^{5/2}}{{\dot m c}\over\alpha\kappa a{\cal R}}, 
\eeq
or
\beq T\approx 10^8 r_6^{-1/5}( r/r_{\rm g})^{3/10}\dot m^{1/5}, \eeq
where $r=10^6r_6$ and $r_{\rm g}=2GM/c^2$ is the gravitational radius of the
accreting object, and the electron scattering opacity of 0.3 cm$^2$/g has been
assumed. The temperatures expected in radiation supported advection dominated
flows are therefore quite low compared with the virial temperature [If the
viscosity is assumed to scale with the total pressure instead of $P_{\rm g}$, the
temperature is even lower]. The effect of electron-positron pairs can be neglected
(Schultz and Price, 1985), since they are present only at temperatures approaching
the electron rest mass energy, $T\gapprox 10^9$K. 

In order for the flow to be radiation pressure and advection dominated, the
optical depth has to be sufficiently large so the radiation does not leak out. The
energy density in the flow, vertically integrated is of the order
\beq E\approx aT^4H,\eeq
and the energy loss rate per cm$^2$ of disk surface is given by (\ref{flux}).
The cooling time is therefore,
\beq t_{\rm c}=E/F=3\tau H/c.\eeq
This is to be compared with the accretion time, which can be written in terms 
of the mass in the disk at radius $r$, of the order $2\pi r^2\Sigma$, and the 
accretion time:
\beq t_{\rm acc}\approx 2\pi r^2\Sigma/\dot M.\eeq
This yields
\beq 
t_{\rm c}/t_{\rm acc}\approx{\kappa\over\pi rc}\dot M={4\over\eta}\dot m{R\over
r}, 
\eeq
(where a factor $3/2\,H/r\sim O(1)$ has been neglected). Since $r>R$, this shows 
that accretion has to be around the Eddington rate or larger in order to be
both radiation- and advection-dominated. 

This condition can also be expressed in terms of the so-called {\em trapping 
radius} $r_{\rm t}$ (e.g. Rees 1978). Equating $t_{\rm acc}$ and $t_{\rm c}$
yields
\beq r_{\rm t}/ R\approx 4 \dot m. \eeq
Inside $r_{\rm t}$, the flow is advection dominated: the radiation field 
produced by viscous dissipation stays trapped inside the flow, instead of being 
radiated from the disk as happens in a standard thin disk. Outside the trapping 
radius, the radiation field cannot be sufficiently strong to maintain a disk 
with $H/r\sim 1$, it must be a thin form of disk instead. Such a thin disk can 
still be radiation-supported (i.e. $P_{\rm r}\gg P_{\rm g}$), but it can not be 
advection dominated.

Flows of this kind are called `radiation supported accretion tori' (or radiation
tori, for short) by Rees et al. 1982. They must accrete at a rate above the
Eddington value to exist. The converse is not quite true: a flow accreting above
Eddington is an advection dominated flow, but it need not necessarily be radiation
dominated.  Advection dominated optically thick accretion flows exist in which
radiation does not play a major role (see section \ref{planet}).

That an accretion flow above $\dot 
M_{\rm E}$ is advection dominated, not a thin disk, also follows from the fact that 
in a thin disk the energy dissipated must be radiated away locally. Since the 
local radiative flux can not exceed the Eddington energy flux $F_{\rm E}$, the 
mass accretion rate in a thin disk can not significantly exceed the Eddington
value (\ref{me}).

The gravitational energy, dissipated by viscous stress in differential rotation
and advected with the flow, ends up on the central object. If this is a black
hole, the photons, particles and their thermal energy are conveniently swallowed
at the horizon, and do not react back on the flow. Radiation tori are therefore
mostly relevant for accretion onto black holes. They are convectively unstable
(Bisnovatyi-Kogan and Blinnikov 1977): the way in which energy is dissipated, in
the standard $\alpha$-prescription, is such that the entropy (entropy of radiation, 
$\sim T^3/\rho$) decreases with height in the disk. Numerical simulations (see 
section \ref{outflows}) show the effects of this convection.

\subsection{Super-Eddington accretion onto black holes}
As the accretion rate onto a black hole is increased above $\dot M_{\rm E}$, the 
trapping radius moves out. The total luminosity increases only slowly, and 
remains of the order of the Eddington luminosity. Such supercritical accretion 
has been considered by Begelman and Meier (1982, see also Wang and Zhou 
1999); they show that the flow has a radially self-similar structure.

Abramowicz et al. (1988, 1989) studied accretion onto black holes at rates near
$\dot M_{\rm E}$. They used a vertically-integrated approximation for the disk,
but included the advection terms. The resulting models were called `slim
disks'. They show how with increasing accretion rate, a standard thin
Shakura-Sunyaev disk turns into a radiation-supported advection flow. The nature
of the transition depends on the viscosity prescription used, and can show a
non-monotonic dependence of $\dot M$ on surface density $\Sigma$ (Honma et al.
1991). This suggests the possibility of instability and cyclic behavior of the
inner disk near a black hole,  at accretion rates near and above $\dot M_{\rm E}$
(for an application to GRS 1915+105 see Nayakshin et al., 1999).

\subsection{Super-Eddington accretion onto neutron stars}
In the case of accretion onto a neutron star, the energy trapped in the flow, plus
the remaining orbital energy, settles onto its surface. If the accretion rate is
below $\dot M_{\rm E}$, the energy can be radiated away by the surface, and steady
accretion is possible. A secondary star providing the mass may, under some
circumstances, transfer more than $\dot M_{\rm E}$, since it does not know about
the neutron star's Eddington value. The outcome of this case is still somewhat
uncertain; it is generally believed on intuitive grounds that the `surplus' (the
amount above $\dot M_{\rm E}$) somehow gets expelled from the system.

One possibility is that, as the transfer rate is increased, the accreting hot gas
forms an extended atmosphere around the neutron star like the envelope of a giant.
If it is large enough, the outer parts of this envelope are partially ionized. The opacity 
in these layers, due to atomic transitions of the CNO and heavier elements, is then
much higher than the electron scattering opacity. The Eddington luminosity based
on the local value of the opacity is then smaller than it is near the neutron star
surface. Once an extended atmosphere forms, the accretion
luminosity is thus large  enough to drive a wind from the envelope (see Kato 1997,
where the importance of this effect is demonstrated in the context of Novae).

This scenario is somewhat dubious however, since it assumes that the mass
transferred from the secondary continues to reach the neutron star and generate a
high luminosity there. This is not at all obvious, since the mass transferring
stream may instead dissipate inside the growing envelope of the neutron star. The
result of this could be a giant (more precisely, a Thorne-Zytkow star), with a
steadily increasing envelope mass. Such an envelope is likely to be large enough
to engulf the entire binary system, which then develops into a common-envelope
(CE) system. The envelope mass is then expected to be ejected by CE hydrodynamics
(for reviews see Taam 1994, Taam and Sandquist 2000).

A more speculative proposal, suggested by the properties of SS 433, is that the
`surplus mass' is ejected in the form of jets. The binary parameters of Cyg X-2
are observational evidence for mass ejection in super-Eddington mass transfer
phases (King and Ritter 1999, Rappaport and Podsiadlowski 2000, King and Begelman
1999).

\section{ADAF Hydrodynamics}
\label{hydro}
The hydrodynamics of radiatively inefficient flows (or `advection dominated accretion 
flows') can be studied by starting, at
a very simple level, with a generalization of the thin disk equations. Making the
assumption that quantities integrated over the height $z$ of the disk give a fair
representation (though this is justifiable only for thin disks), and assuming
axisymmetry, the problem reduces to a one-dimensional time-dependent one. 
Further simplifying this by restriction to a steady flow yields the equations
\beq 2\pi r\Sigma v_r=\dot M={\rm cst} \label{cnt},\eeq
\beq 
r\Sigma v_r\p_r(\Om r^2)=\p_r(\nu\Sigma r^3\p_r\Om),\label{ang}
\eeq
\beq 
v_r\p_r v_r-(\Om^2-\Om^2_{\rm K})r=-{1\over\rho}\p_r P,\label{vr}
\eeq 
\beq\Sigma v_rT\p_rS=q^+-q^-, \label{eng}\eeq
where $S$ is the specific entropy of the gas,  $\Om$ the local rotation rate, 
now different from the Keplerian rate $\Omk$, and 
\beq 
q^+=\int Q_{\rm v}\rmd z  \qquad q^-=\int {\rm div}F_{\rm r}\rmd z \label{qs}
\eeq
are the height-integrated viscous dissipation rate and radiative loss rate, 
respectively. In the case of thin disks, equations (\ref{cnt}) and (\ref{ang}) are 
unchanged, but (\ref{vr}) simplifies to $\Om^2=\Omk^2$, i.e. the rotation is 
Keplerian, while (\ref{eng}) simplifies to $q^+=q^-$, expressing local balance 
between viscous dissipation and cooling. The left hand side of (\ref{eng}) 
describes the radial advection of heat, and is perhaps the most important 
deviation from the thin disk equations at this level of approximation (hence the 
name advection dominated flows). The characteristic properties are seen most 
clearly when radiative loss is neglected altogether, $q^-=0$. 
The equations are supplemented with expressions for $\nu$ and $q^+$:
\beq 
\nu=\alpha c_{\rm s}^2/\Omk; \qquad q^+=(r\p_r\Om)^2\nu\Sigma.
\eeq
If $\alpha$ is taken constant, $q^-=0$, and an ideal gas is assumed with 
constant ratio of specific heats, so that the entropy is given by
\beq S=c_{\rm v}\ln (p/\rho^\gamma), \eeq
then equations (\ref{cnt})-(\ref{eng}) have no explicit length scale in them. This 
means that a special so-called self-similar solution exists, in which all 
quantities are powers of $r$.  Such self-similar solutions have apparently been 
described first by Gilham (1981), but re-invented several times (Spruit et al. 1987; 
Narayan and Yi, 1994). The dependences on $r$  are 
\beq \Om\sim r^{-3/2};\qquad \rho\sim r^{-3/2},\label{sc1}\eeq
\beq H\sim r; \qquad T\sim r^{-1}.\label{sc2}\eeq
In the limit $\alpha\ll 1$, one finds
\beq 
v_r=-\alpha\Omk r \left(9{\gamma-1\over 5-\gamma}\right),
\eeq
\beq 
\Om=\Omega_{\rm K}\left(2{5-3\gamma\over 5-\gamma}\right)^{1/2},
\eeq
\beq c_{\rm s}^2=\Omk^2r^2{\gamma-1\over 5-\gamma}, \eeq
\beq {H\over r}=\left(\gamma-1\over 5-\gamma\right)^{1/2}.\eeq
The precise from of these expressions depends somewhat on the way in which
vertical integrations such as in (\ref{qs}) are done (which are only approximate).

The self-similar solution can be compared with numerical solutions of eqs.
(\ref{cnt})--(\ref{eng}) with appropriate conditions applied at inner ($r_{\rm
i}$) and outer ($r_{\rm o}$) boundaries (Nakamura et al. 1996, Narayan et al.
1997). The results show that the self-similar solution is valid in an intermediate
regime $r_{\rm i}\ll r\ll r_{\rm o}$. That is, the solutions of
(\ref{cnt})--(\ref{eng}) approach the self-similar solution far from the
boundaries, as is characteristic of self-similar solutions.

The solution exists only if $1<\gamma\le 5/3$, a condition satisfied by all ideal
gases. As $\gamma\downarrow 1$, the disk temperature and thickness vanish. This is
understandable, since a $\gamma$ close to 1 means that the particles making up the
gas have a large number of internal degrees of freedom. In thermal
equilibrium the accretion energy is shared between all degrees of freedom, so that
for a low $\gamma$ less is available for the kinetic energy (temperature) of the
particles.

Second, the {\em rotation rate vanishes} for $\gamma\rightarrow 5/3$. As in the
case of spherical accretion no accreting solutions exist for $\gamma>5/3$ (cf.
section \ref{bondi}). Since a fully ionized gas has $\gamma=5/3$, it is a
relevant value for the hot, ion supported accretion flows discussed below.
Apparently, steady advection dominated accretion can not have angular momentum in
this case. To see how comes about consider the entropy of the flow. For accretion to 
take place in a rotating flow, there has to be friction and increase of entropy. Accretion 
is then necessarily accompanied by an inward increase of entropy. In a
radially self similar flow, the scalings (\ref{sc1}), (\ref{sc2}) yield $P\sim r^{-5/2}$,
so entropy scales as $S\sim \ln P/\rho^\gamma\sim (5-3\gamma)\ln r$. This increases
with decreasing distance $r$ only for $\gamma<5/3$, and an accretion flow with 
$\gamma=5/3$ cannot be both rotating and adiabatic. (With energy losses by
radiation the constraint on $\gamma$ disappears again).

The question then arises how an adiabatic flow with  $\gamma=5/3$ will
behave if one starts it as a rotating torus around a black hole. In the
literature, this problem has been circumvented by arguing that real flows would
have magnetic fields in them, which would change the effective compressibility of
the gas. Even if a magnetic field of sufficient strength is present, however,
(energy density comparable to the gas pressure) the effective $\gamma$ is not
automatically lowered. If the field is compressed mainly perpendicular to the
field lines, for example, the effective $\gamma$ is closer to 2. Also, this does
not solve the conceptual problem what would happen to a rotating accretion flow
consisting of a more weakly magnetized ionized gas.

\subsubsection{The case of the vanishing rotation rate}
This conceptual problem has been solved by Ogilvie (1999), who showed how a gas
cloud initially rotating around a point mass settles to the slowly rotating
self-similar solutions of the steady problem discussed above. He constructed
similarity solutions to the time dependent version of eqs (\ref{cnt})--(\ref{eng}),
in which distance and time occur in the combination $r/t^{2/3}$. This solution
describes the asymptotic behavior (in time) of a viscously spreading disk,
analogous to the viscous spreading of thin disks (see section \ref{spread}). As in
the thin disk case, all the mass accretes asymptotically onto the central mass,
while all the angular momentum travels to infinity together with a vanishing
amount of mass. For all $\gamma<5/3$, the rotation rate at a fixed $r$ tends to a
finite value as $t\rightarrow\infty$, but for $\gamma=5/3$ it tends to zero. The
size of the slowly-rotating region expands as $r\sim t^{2/3}$.The typical slow 
rotation of ADAFs at $\gamma$ near 5/3 is thus a real
physical property. In such a flow the angular momentum gets expelled from the
inner regions almost completely, and the accretion flow becomes purely radial,
as in Bondi accretion.

\subsection{Other optically thick accretion flows}
\label{planet}
The radiation-dominated flows discussed in section \ref{radv} are not the only
possible optically thick advection dominated flows. From the discussion of the
hydrodynamics, it is clear that disk-like (i.e. rotating) accretion is possible
whenever the ratio of specific heats is less than 5/3. A radiation supported flow
satisfies this requirement since radiation pressure scales with volume like a 
gas with $\gamma=4/3$, but it can also happen in the
absence of radiation if energy is taken up in the gas by internal degrees of
freedom of the particles. Examples are the rotational and vibrational degrees of
freedom in molecules, and the energy associated with dissociation and ionization.
If the accreting object has a gravitational potential not too far from the 2.3 +
13.6 eV per proton for dissociation plus ionization, a gas initially consisting of
molecular  hydrogen can stay bound at arbitrary accretion rates. This translates
into a limit $M/M_\odot\, R_\odot/R<0.01$. This is satisfied approximately by the
giant planets, which are believed to have gone through a phase of rapid adiabatic
gas accretion  (e.g. Podolak et al. 1993).

A more remotely related example is the core-collapse supernova. The accretion 
energy of the core mass falling onto the growing proto-neutron star at its
center is lost mostly 
to internal degrees of freedom represented by photodisintegration of the nuclei. 
If the pre-collapse core rotates sufficiently rapidly, the collapse will form an 
accretion  torus (inside the supernova envelope), with properties similar to 
advection dominated accretion flows (but at extreme densities and accretion 
rates, by X-ray binary standards). Such objects have been invoked as sources of 
Gamma-ray bursts (Woosley 1993, Paczy\'nski 1998, Popham et al. 1999).

A final possibility for optically thick accretion is through {\em neutrino 
losses}. If the temperature and density near an accreting neutron star become  
large enough, additional cooling takes place through neutrinos (as in the cores 
of giants). This is relevant for the physics of Thorne-Zytkow stars (neutron 
stars or black holes in massive supergiant envelopes, cf.\ Bisnovatyi-Kogan and 
Lamzin 1984, Cannon et al. 1992), and perhaps for the spiral-in of neutron stars 
into giants (Chevalier 1993, see however Taam \& Sandquist 2000).

\section{Optically thin radiatively inefficient flows (ISAFs)}
\label{isaf}
The optically thin case has received most attention, because of the promise it
holds for explaining the (radio to X-ray) spectra of X-ray binaries and the
central black holes in galaxies, including our own.  For a review see Yi (1999).
This kind of flow occurs if the gas is optically thin, and radiation processes
are sufficiently weak. The gas then heats up to near the virial temperature. Near the
last stable orbit of a black hole, this is of the order 100 MeV, or $10^{12}$ K.
At such temperatures, a gas in thermal equilibrium would radiate at a fantastic
rate, even if it were optically thin, because the interaction between electrons
and photons becomes very strong already near the electron rest mass of 0.5 MeV. In
a remarkable early paper, Shapiro, Lightman and Eardley (1976) noted that this,
however, is not what will happen in an optically thin plasma accreting on a hole. They
showed that instead thermal equilibrium between ions and electrons breaks down
and a {\it two-temperature plasma} forms. We call such a flow an ion supported
accretion flow (ISAF), following the nomenclature suggested by Rees et al. (1982).
The argument is as follows.

Suppose that the energy released by viscous dissipation is distributed equally
among the carriers of mass, i.e. mostly to the ions and $\sim 1/2000$ to the
electrons. Most of the energy then resides in the ions, which radiate very
inefficiently (their high mass prevents the rapid accelerations that are needed to
produce electromagnetic radiation). Their energy is transfered to the electrons by
Coulomb interactions. These interactions are slow, however, under the conditions
mentioned. They are slow because of the low density (on account of the assumed
low optical thickness), and because they decrease with increasing temperature. The
electric forces that transfer energy from an ion to an electron act only as long
as the ion is within the electron's Debye sphere (e.g.\ Spitzer, 1965). The
interaction time between proton and electron, and thus the momentum transfered,
therefore decrease as $1/v_{\rm p}\sim T_{\rm p}^{-1/2}$ where $T_{\rm p}$ is the
proton temperature.

In this way, an optically thin plasma near a compact object can be in a
two-temperature state, with the ions being near the virial temperature and the
electrons, which are doing the radiating, at a much lower temperature around
50--200 keV. The energy transfer from the gravitational field to the ions is fast
(by some form of viscous or magnetic dissipation, say), from the ions to the
electrons slow, and finally the energy losses of the electrons fast (by
synchrotron radiation in a magnetic field or by inverse Compton scattering off soft
photons). Such a flow would be radiatively inefficient since the receivers of the
accretion energy, the ions, get swallowed by the hole before having a chance to
transfer their energy to the electrons. Most of the accretion energy thus gets
lost into the hole, and the radiative efficiency is much less than for a cool disk. 
The first disk models which take into account this physics of advection and 
a two-temperature plasma were developed by Ichimaru (1977).

\begin{figure}
\hfil\includegraphics[width=8 cm, clip]{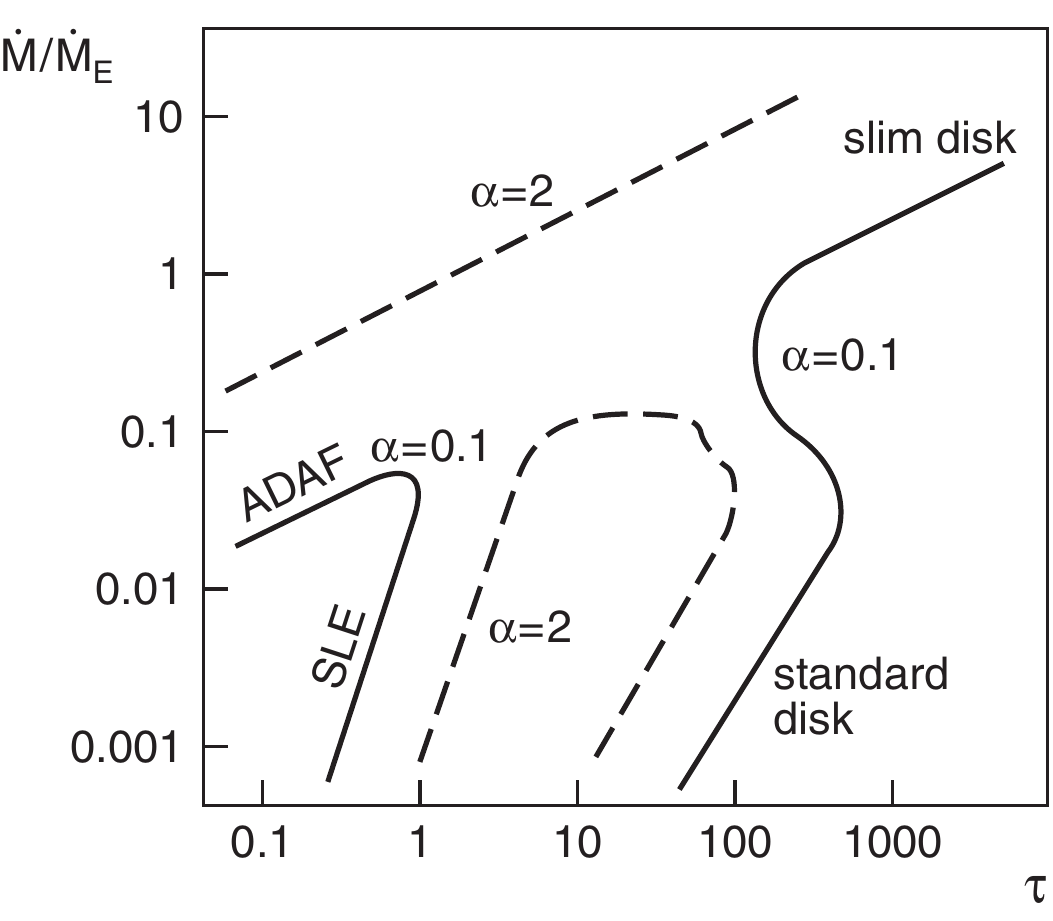}\hfil
\caption{Branches of advection-dominated and thin disks for two values or the 
viscosity parameter $\alpha$, as functions of accretion rate and (vertical)
optical depth of the flow (schematic, after Chen et al. 1995, Zdziarski 1998).
Optically thin branches are the ISAF and SLE (Shapiro-Lightman-Eardley) solutions,
optically thick ones the radiation dominated (`slim disk' or `radiation torus')
and SS (Shakura-Sunyaev or standard thin disk). Advection dominated are the ISAF
and the radiation torus, geometrically thin are the SLE and SS. The SLE solution
is a thermally unstable branch.}\label{branch}
\end{figure}

It is clear from this description that both the physics of such flows and the
radiation spectrum to be expected depend critically on the details of the
ion-electron interaction and radiation processes assumed. This is unlike the case
of the optically thick advection dominated flows, where gas and radiation are in
approximate thermodynamic equilibrium. This is a source of uncertainty in the
application of ISAFs to observed systems, since their radiative properties depend
on poorly known quantities such as the strength of the magnetic field in the flow.

The various branches of optically thin and thick accretion flows are summarized in
figure \ref{branch}. Each defines a relation between surface density $\Sigma$ (or optical
depth $\tau=\kappa\Sigma$) and accretion rate. ISAFs require low densities, which
can result either because of low accretion rates, or large values of the viscosity
parameter. The condition that the cooling time of the ions by energy transfer to
the electrons is longer than the accretion time yields a maximum accretion rate
(Rees et al. 1982), \beq \dot m\lapprox \alpha^2, \eeq 
where $\dot m$ is the accretion rate in units of the Eddington value.
If $\alpha\approx 0.05$ as
suggested by current simulations of magnetic turbulence, the maximum accretion
rate would be a few $10^{-3}$. If ISAFs are to be applicable to systems with
higher accretion rates, such as Cyg X-1 for example, the viscosity parameter must
be larger, on the order of 0.3.

\subsection{Application: hard spectra in X-ray binaries}
In the hard state, the X-ray spectrum of black hole and neutron star accreters is
characterized by a peak in the energy distribution ($\nu F_\nu$ or $E\,F(E)$) at
photon energies around 100 keV. This is to be compared with the typical photon
energy of $\sim 1$ keV expected from a standard optically thick thin disk
accreting near the Eddington limit. The standard, and by far most likely
explanation is that the observed hard photons are softer photons (around 1 keV)
that have been up-scattered through inverse Compton scattering on hot electrons.
Fits of such Comptonized spectra (e.g. Sunyaev and Titarchuk 1980, Zdziarski 1998
and references therein) yield an electron scattering optical depth around unity
and an electron temperature of 50--100 keV. The scatter in these parameters is
rather small between different sources. The reason may lie in part in the physics
of Comptonization, but this is not the only reason. Something in the physics of
the accretion flow keeps the Comptonization parameters fairly constant as long as it is
in the hard state. ISAFs have been applied with some success in interpreting XRB.
They can produce reasonable X-ray spectra, and have been used in interpretations
of the spectral-state transitions in sources like Cyg X-1 (Esin et al. 1998 and
references therein).

An alternative to the ISAF model for the hard state in sources like Cyg X-1 and 
the black hole X-ray transients is the `corona' model. A hot corona 
(Bisnovatyi-Kogan and Blinnikov 1976), heated perhaps by magnetic fields as in 
the case of the Sun  (Galeev et al. 1979) could be the medium that Comptonizes 
soft photons radiated from the cool disk underneath. The energy balance in such 
a model produces a Comptonized spectrum within the observed range (Haardt 
and Maraschi 1993). This model has received further momentum,  especially as 
a model for AGN, with the discovery of broadened X-ray lines interpreted as 
indicative of the presence of a cool disk close to the last stable orbit around 
a black hole (Fabian et al. 2002 and references therein). The very rapid X-ray 
variability seen in some of these sources is interpreted as due magnetic flaring 
in the corona, like in the solar corona (e.g. Di Matteo et al. 1999a).

\subsection{Transition from cool disk to ISAF: truncated disks}
One of the difficulties in applying ISAFs to specific observed systems is the 
transition from a standard geometrically thin, optically thick disk, which must 
be the mode of mass transfer at large distances, to an ISAF at closer range. 
This is shown by figure \ref{branch}, which illustrates the situation at some distance 
close to the central object. The standard disk and the optically thin branches 
are separated from each other for all values of the viscosity parameter. This 
separation of the optically thin solutions also holds at larger distances. Thus, 
there is no plausible continuous path from one to the other, and the transition 
between the two must be due to additional physics that is not included in 
diagrams like figure \ref{branch}. 

Circumstantial observational evidence points to the existence of such a transition. The 
distance from the hole where it is assumed to take place is then called the truncation 
radius. The extensive datasets from the black hole candidate Cyg X-1  obtained with 
the Rossi X-ray Timing Explorer (RXTE) have played an important role in the 
development of the truncated disk model. The X-ray spectrum of Cyg X-1 varies (on 
time scales of days to years) between softer and harder states, and the characteristic 
time scale of its fast variability (milliseconds to minutes) correlates 
closely  with these changes. Since all disk time scales decrease with distance from 
the hole, the fast variability is interpreted as due to a process (still to be identified in 
detail) that depends on the size of the truncation radius. Variation in time of this radius 
is then assumed to cause the observed changes. Characteristics of the spectrum 
that find a natural place in this picture are: the 
slope of the hard X-ray spectrum, the amplitude of the so-called `Compton reflection 
hump', and the behavior of the Fe K$_\alpha$ fluorescence line at 6.7 keV; each of 
these  correlates with the characteristic fast variability time scale in an interpretable 
(though still somewhat model-dependent) way (Gilfanov et al. 1999, Revnivtsev et al. 
1999).  Very similar correlations have been found in X-ray observations of AGN
(Zdziarski et al. 1999).

A promising possibility is that the transition takes place through {\em
evaporation}. Two distinct mechanisms have been elaborated for such evaporation.
In the first (Meyer and Meyer-Hofmeister 1994, Liu et al. 2002), the evaporation
starts at a relatively large distance from the hole, where the virial temperature
is of the order of $10^6-10^7$ K. As in the solar corona, the strong decrease of
radiative efficiency of gas with temperature in this range produces a hot
optically thin corona in contact with the cool disk below, and exchange of mass
can take place through evaporation and condensation, and the process is mediated
by electron heat conduction. In this scenario, a corona flow at $\sim 10^7$ K at a
distance of several hundred Schwarzschild radii transforms into a two-temperature
ISAF further in. 

\subsubsection{Ion illumination}
\begin{figure}
 \includegraphics[width=12 cm, clip]{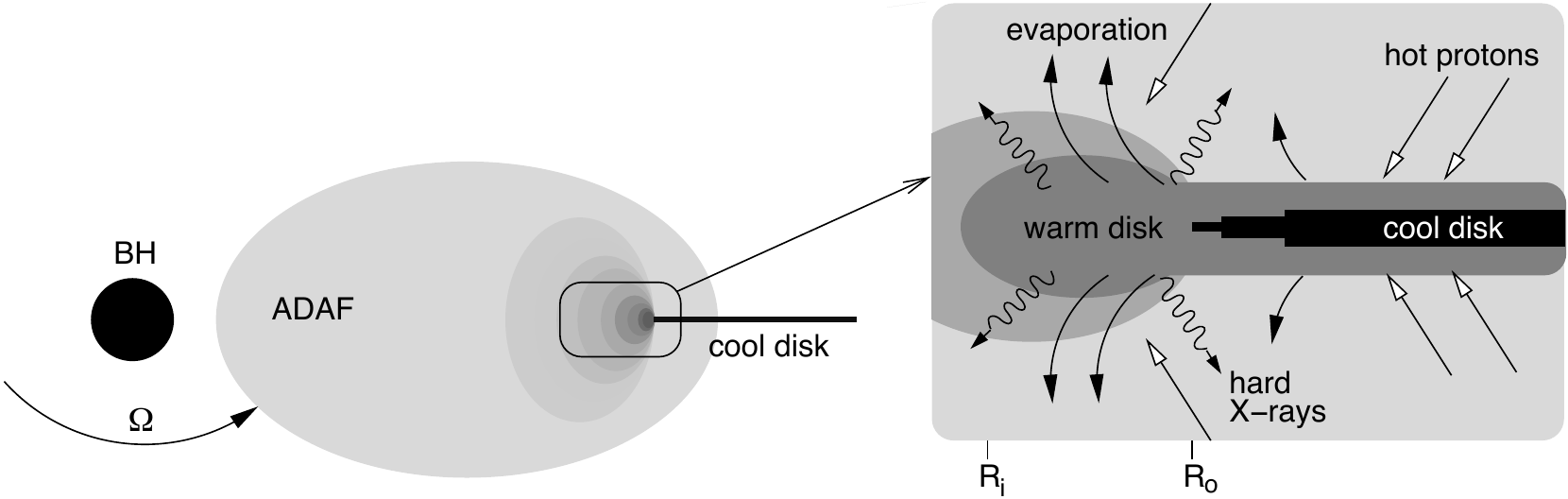}\hfil
\caption[]{\label{ionillum}
interaction between ion supported accretion flow(ISAF) and cool disk in the 
truncated disk model for hard X-ray states in X-ray binaries. Accreting black hole 
on the left, its hot ions illuminating the cool disk create a warm ($\sim 100$ keV 
surface layer producing the hard X-rays, extending inward and heating up to virial 
temperature by thermal instability  (`evaporation') once cooling by soft photons 
becomes inefficient}   
\end{figure} 

Observations indicate that cool disks can also coexist with a hot, hard X-ray
producing plasma quite close to the hole (for references see Dullemond and Spruit
2004). At these close distances, evaporation must behave differently from the
coronal evaporation model, since the interaction of a two-temperature plasma with
a cool disk is very different from that of a plasma at coronal temperatures
(Spruit 1997). The energy in an ISAF is in the ions, and the electron conduction of
heat that drives coronal evaporation unimportant. Moreover, the ions penetrate a 
substantial distance into the cool disk, and the energy they dump is radiated away 
long before they can heat up the disk to virial
temperatures. Nevertheless, evaporation can still take place in this case, since
it turns out that the interaction of the ions with the cool disk produces a layer
of intermediate temperature (around 100 keV) that becomes thermally unstable in
the presence of viscous dissipation, and heats up ISAF temperatures (Deufel et al.
2000, 2001, 2002, see also Spruit and Deufel 2002). The sequence of events
is illustrated in figure \ref{ionillum}. This model explains both the
hard spectra of typical black hole accreters and the coexistence of cool and hot
plasma that is indicated by the observations (Dullemond and Spruit 2004, D'Angelo
et al. 2008).

\subsection{Quiescent galactic nuclei}
For very low accretion rates, such as inferred for the black hole in the center of
our galaxy (identified with the radio source Sgr A*), the broad band spectral
energy distribution of an ISAF is predicted to have two humps (Narayan et al.
1995, Quataert et al. 1999). In the X-ray range, the emission is due to
bremsstrahlung. In the radio range, the flow emits synchrotron radiation, provided
that the magnetic field in the flow has an energy density order of the gas
pressure (`equipartition'). Synthetic ISAF spectra can be fitted to the observed
radio and X-ray emission from Sgr A*. In other galaxies where a massive central
black hole is inferred, and the center is populated by an X-ray emitting gas of
known density, ISAFs would also be natural, and might explain why the observed
luminosities are so low compared with the accretion rate expected for a hole
embedded in a  gas of the measured density.

\subsection{ISAF-disk interaction: Lithium}
One of the strong predictions of ISAF models, whether for black holes or neutron
stars, is that the accreting plasma in the inner regions has an ion temperature of
10--100 MeV. Nearby is a cool and dense accretion disk feeding this plasma. If
only a small fraction of the hot ion plasma gets in contact with the disk, the
intense irradiation by ions will produce nuclear reactions (Aharonian and Sunyaev
1984, Mart\'{\i}n et al. 1992). The main effects would be spallation of CNO
elements into Li and Be, and the release of neutrons by various reactions. In this
context, it is intriguing that the secondaries of both neutron star and black hole
accreters have high overabundances of Li compared with other stars of their
spectral types (Mart\'{\i}n et al. 1992, 1994a). If a fraction of the disk
material is carried to the secondary by a disk wind, the observed Li abundances
may be accounted for (Mart\'{\i}n et al. 1994b).

\section{Outflows}
\label{outflows}
The energy density in an advection dominated accretion  flow is of the same 
order as the gravitational binding energy density $GM/r$, since a significant 
fraction of that energy went into internal energy of the gas by viscous 
dissipation, and little of it got lost by radiation. The gas is therefore only 
marginally bound in the gravitational potential. This suggests that perhaps a 
part of the accreting gas can escape, producing an outflow or wind. In the case 
of the ion supported ISAFs, this wind would be thermally driven 
by the temperature of the ions, like an `evaporation' from the accretion torus. In
the case of the radiation supported tori, which exist only at a luminosity near
the Eddington value, but with much lower temperatures than the ion tori, winds
driven by radiation pressure could exist.

The possibility of outflows is enhanced by the viscous energy transport through 
the disk. In the case of thin accretion disks (not quite appropriate in the
present case, but sufficient to demonstrate the effect), the local rate of
gravitational energy release (erg cm$^{-2}$s$^{-1}$) is $W=\Sigma v_r \p_r(GM/r)$.
The local viscous dissipation rate is $(9/4)\nu\Sigma\Om^2$. As discussed in 
section \ref{visgrav} they are related by 
\beq Q_{\rm v}=3[1-({r_{\rm i}\over r})^{1/2}]W, \eeq
where $r_{\rm i}$ is the inner edge of the disk. Part of the gravitational energy 
released in the inner disk is transported
outward by the viscous stresses, so that the  energy deposited in the gas is up to
three times larger than expected from a local energy balance argument. The
temperatures in an ADAF would be correspondingly larger. Begelman and Blandford
(1999) have appealed to this effect to argue that in an ADAF most of the accreting
mass of a disk might be expelled through a wind, the energy needed for this being
supplied by the viscous energy transport associated with the small amount of mass
that actually accretes.

Most outflows from disks such as the observed relativistic jets (and the non-relativistic 
ones in protostellar objects) are now believed to have a magnetic origin, requiring the 
presence of a strong, ordered (large scale) magnetic field anchored in the disk. Three-
dimensional numerical MHD simulations of such processes are becoming increasingly 
realistic (e.g.\ Krolik \& Hawley 2010, Moll 2009). The origin of jet-friendly magnetic field 
configurations, on tge other hand, and the reasons for their apparently unpredictable 
presence are still unknown (not all black holes show jets, and the ones that do don't 
have them all the time). There is, however, a degree of correlation of their presence 
with  the hard X-ray states in these objects (in the sense that they appear absent in soft 
states). For a discussion and interpretation of these issues see Spruit (2008).

\bigskip
\medskip
\leftline{\bf \large References}
\medskip


\rea Abramowicz, M.A., Czerny B., Lasota J.-P., Szuszkiewicz E., 1988, \apj{332}, 646

\rea Abramowicz M.A, Kato S., Matsumoto R., 1989, \pasj{41}, 1215

\rea Aharonian F.A., Sunyaev, R.A., 1984 \mnras{210}, 257

\rea  Armitage P.J., 1998, \apj{501}, L189

\rea  Armitage P.J., and Livio M., 1998, \apj{493}, 898

\rea  Balbus S.A. and Hawley J.F., 1991, \apj{376}, 214

\rea  Balbus, S.~A.\ 2003, \araa{41}, 555 

\rea  Begelman M.C., 1979, \mnras{187}, 237

\rea Begelman M.C., Meier D.L., 1982, \apj{253}, 873 

\rea  Begelman M.C., McKee C.F., Shields G.A., 1983, \apj{271}, 70

\rea Begelman M.C., Blandford R.D., 1999, \mnras{303}, L1

\rea  Bisnovatyi-Kogan G., 1993, \aaa{274}, 796

\rea  Bisnovatyi-Kogan G., and Ruzmaikin A.A. 1976, \ass{42}, 401

\rea Bisnovatyi-Kogan G.S., Blinnikov S.I., 1977, \aaa{59}, 111

\rea Bisnovatyi-Kogan G.S., Lamzin S.A., 1984, {\it Astron. Zh} {\bf 61}, 323 (translation {\it Soviet Astron.} {\bf 28}, 187)

\rea  Blaes O. and Hawley J.F., 1988, \apj{326}, 277

\rea  Blandford R.D., 1976, \mnras{176}, 465

\rea  Blandford R.D. and Payne D.G., 1982 \mnras{199}, 883

\rea  Bondi H., 1952, \mnras{112}, 195

\rea  Bradshaw P., 1969, {\it J. Fluid Mech.} {\bf 36}, 177

\rea  Brandenburg A., Nordlund {\AA}., Stein R.F. and Torkelsson U., 1995, \apj{446}, 741

\rea Cannon R.C., Eggleton P.P., Zytkow A.N., Podsiadlowski P., 1992,
\apj{386}, 206

\rea  Cannizzo J.K., Shafter A.W., Wheeler J.C., 1988, \apj{333},227

\rea  Chen X.M., Abramowicz M.A., Lasota J.-P., Narayan R., Yi I., 1995, \apj{443}, L61

\rea Chevalier R.A., 1993, \apj{411}, 33

\rea  D'Angelo, C., Giannios, D., Dullemond, C., \& Spruit, H.\ 2008, \aap{488}, 441 

\rea  D'Angelo, C.~R., \& Spruit, H.~C.\ 2010, \aap in press, arXiv:1001.1742 

\rea  Cordova F.A., 1995, in W.H.G. Lewin et al. eds. {\it X-ray Binaries},
Cambridge University Press, Cambridge, p.331

\rea  Courant R. and Friedrichs K.O., 1948, {\it Supersonic Flow and Shock
Waves}, Interscience, New York, and Springer, Berlin (1976)

\rea  Chandrasekhar C., 1961, {\it Hydrodynamic and Hydromagnetic
Stability}, Oxford Univ. Press, Oxford, and Dover Publications Inc., 1981

\rea  Davis, S.~W., Stone, J.~M., \& Pessah, M.~E.\ 2010, \apj{713}, 52

\rea  De Villiers, J., \& Hawley, J.F.\ 2003, \apj{592} 1060

\rea {}Di Matteo T., Celotti A., Fabian A.C., 1999a, \mnras{304}, 809

\rea  Di Matteo T., Fabian A. C., Rees M.J., Carilli C.L., Ivison
R.J.,1999b,  \mnras{305}, 492

\rea  Deufel, B., \& Spruit, H.~C.\ 2000, \aap{362}, 1 

\rea  Deufel, B., Dullemond, C.P., \& Spruit, H.C.\ 2001, \aap{377}, 955 

\rea  Deufel, B., Dullemond, C.P., \& Spruit, H.C.\ 2002, \aap{387}, 907 

\rea  Dubrulle, B.\ 1992, \aap{266}, 592 

\rea  Dullemond, C.P. \& Spruit, H.C., 2005, \aap{434}, 415

\rea  Dullemond, C.~P., Dominik, C., \& Natta, A.\ 2001, \apj{560}, 957 

\rea  Eardley D., and Lightman A., 1975, \apj{200}, 187

\rea  Emmering R.T., Blandford R.D., Shlosman I., 1992, \apj{385}, 460

\rea  Esin A.A., Narayan R., Cui W., Grove J.E., Zhang S.N., 1998,
\apj{505}, 854

\rea  Fabian, A.C., et al.\ 2002, \mnras{335}, L1 

\rea  Frank J., King A.R., and Raine D.J., 2002, {\it Accretion Power in
Astrophysics} (3rd edition), Cambridge University Press.

\rea  S.~Fromang, J.C.B.~Papaloizou, G.~Lesur et al. 2007a, \aap{476}, 1123

\rea  Fromang, S., Papaloizou, J., Lesur, G., \& Heinemann, T.\ 2007b, \aap{476}, 1123 

\rea Galeev A.A., Rosner R., Vaiana G.S., 1979, \apj{229}, 318

\rea Gammie, C.~F.\ 1997, IAU Colloq.~163: Accretion Phenomena and Related Outflows, ed. D. T. Wickramasinghe et al. 121, 704

\rea  Gilfanov, M., Churazov, E., \& Revnivtsev, M.\ 1999, \aap{352}, 182 

\rea  Gilham S., 1981, \mnras{195}, 755

\rea Haardt F., Maraschi L., Ghisellini G., 1994, \apj{432}, 95

\rea Hameury J.-M., Menou K., Dubus G., Lasota J.-P., Hure J.-M., 1998,
\mnras{298}, 1048

\rea  Hawley J.E., 1991 \apj{381}, 496

\rea  Hawley  J.F., Gammie  C.F. and Balbus S.A., 1995, \apj{440}, 742

\rea  Hawley J.F., Balbus S.A., Winters W.F. 1999, \apj{518}, 394

\rea  Hawley, J.~F., Balbus, S.~A., \& Winters, W.~F.\ 1999, \apj{518}, 394 

\rea  Hirose, S., Krolik, J.H., De Villiers, J., \& Hawley, J.F.\ 2004, \apj{606}, 1083 

\rea  Hirose, S., Krolik, J.~H., \& Blaes, O.\ 2009, \apj{691}, 16 

\rea Honma F., Kato S., Matsumoto R., Abramowicz M., 1991, \pasj{43}, 261

\rea  Horne K., 1985, \mnras{213}, 129

\rea  Horne, K., 1993, in J.C. Wheeler, ed., {\it Accretion Disks in
Compact Stellar Systems}, Singapore: World Scientific Publishing

\rea  H\=oshi R., 1979, {\it Prog. Theor. Phys.} {\bf 61}, 1307

\rea  Hurley, K., et al.\ 2005, \nature{434}, 1098 

\rea Ichimaru S., 1977, \apj{241}, 840

\rea Igumenshchev I.V., Chen X.M., Abramowicz A., 1996 \mnras{278}, 236

\rea Igumenshchev I.V., Abramowicz M.A., 1999, \mnras{303}, 309

\rea Illarionov A.F., Sunyaev R.A., 1975, \aaa{39}, 185

\rea  Inogamov N.A., Sunyaev R.A., 1999, Astron. Lett., 1999, 25, 269 
(=astro-ph/9904333)

\rea  Ji, H., Burin, Schartman, M.E. \& J. Goodman, J., 2006, Nature {\bf 444}, 343 

\rea  Kato M., 1997, \apjs{113}, 121

\rea  Kippenhahn R. and Thomas H.-C., 1981, in D. Sugimoto, D. Lamb and
D.N. Schramm, {\it Fundamental Problems in the Theory of Stellar Evolution}
(IAU Symp. {\bf 93}), Reidel, Dordrecht, p.237

\rea  King A.R., 1995 in Lewin, W.H.G., van Paradijs, J., van den Heuvel,
E.P.J., eds., {\it X-ray Binaries}, Cambridge University press

\rea  King, A.~R.\ 2004, \mnras{347}, L18 

\rea  King A.R. and Ritter H., 1999 \mnras{309}, 253

\rea  King A.R., Begelman M.C., 1999, \apj{519}, L169

\rea  K\"onigl A., 1989 \apj{342}, 208

\rea  K\"onigl A.,  Kartje J.F., 1994 \apj{434}, 446

\rea  Klahr, H.~H., \& Bodenheimer, P.\ 2003, \apj{582}, 869 

\rea  Kley W., 1991, \aaa{247}, 95

\rea  Kley,W., Papaloizou J.C.B., Lin D.N.C, 1993, \apj{416}, 679

\rea  Krolik, J.~H., \& Hawley, J.~F.\ 2010, Lecture Notes in Physics, Berlin: Springer  {\bf 794}, 265 

\rea  Kuulkers E., Howell S.B., van Paradijs J., 1996, \apj{462}, 87

\rea  Landau L.D. and Lifshitz E.M., 1959 {\it Fluid Mechanics}, Pergamon Press, Oxford

\rea  Lee, C.-F., Mundy, L.~G., Reipurth, B., Ostriker, E.~C., \& Stone, J.~M.\ 2000, \apj, 542, 925 

\rea  Lesur, G., \& Longaretti, P.-Y.\ 2005, \aap{444}, 25 

\rea  Lesur, G., \& Ogilvie, G.~I.\ 2008, \aap{488}, 451 

\rea  Lesur, G., \& Papaloizou, J.~C.~B.\ 2010, \aap{513}, A60 

\rea  Lewin W.H.G., van Paradijs J., van den Heuvel E.P.J., 1995, {\it X-ray Binaries}, Cambridge University press

\rea  Lightman A.P.,  Eardley D.M., 1974, \apj{187}, L1

\rea  Lin D.N.C., and Taam R.E., 1984, in S.E. Woosley, ed., High Energy
Transients in Astrophysics, AIP Conference Procs. {\bf 115}, p.83

\rea  Liu, B.F., Mineshige, S., Meyer, F., Meyer-Hofmeister, E., \& Kawaguchi, T.\ 2002, \apj{575}, 117

\rea  Long, M., Romanova, M.~M., \& Lovelace, R.~V.~E.\ 2008, \mnras{386}, 1274 

\rea  Lovelace R.V.E., 1976, \nature{262}, 649

\rea  L\"ust R., 1952, {\it Z. Naturforsch.} {\bf 7a}, 87

\rea  Lynden-Bell D. and Pringle J.E., 1974, \mnras{168}, 603 

\rea  Lyutyi V.M. and Sunyaev R.A., 1976, {\it Astron. Zh.} {\bf 53}, 511,
translation in {\it Sov. astron.} {\bf 20}, 290 (1976)

\rea Mart\'{\i}n E L., Rebolo R., Casares J., Charles P.A., 1992, {\it
Nature} {\bf 358}, 129

\rea Mart\'{\i}n E.L., Rebolo R., Casares J., Charles P.A., 1994a
\apj{435}, 791

\rea Mart\'{\i}n E., Spruit H.C., van Paradijs J., 1994b, \aaa{291}, L43

\rea  Massey B.S., 1968, {\it Mechanics of Fluids}, Chapman and Hall, London (6th Ed. 1989)

\rea  Meyer F., 1990, {\it Rev. Mod. Astron.} {\bf 3}, 1

\rea  Meyer F. and Meyer-Hofmeister E., 1981, \aaa{104}, L10

\rea  Meyer F. and Meyer-Hofmeister E., 1982, \aaa{106}, 34

\rea  Meyer-Hofmeister, E. and Ritter, H., 1993, in {\it The Realm of Interacting Binary Stars}, ed. J. Sahade, Kluwer, Dordrecht, p.143

\rea Meyer F., Meyer-Hofmeister E., 1994, \aaa{288}, 175

\rea  Meyer-Hofmeister E., Schandl S., and Meyer F.,1997 \aaa{318}, 73

\rea  Meyer-Hofmeister E., Meyer F., 1999, astro-ph/9906305

\rea  Migliari, S., \& Fender, R.~P.\ 2006, \mnras{366}, 79 

\rea  Miller, K.A., and Stone, J.M., 1997, \apj{489}, 890

\rea  Mineshige S. and Wheeler J.C., 1989 \apj{343}, 241

\rea  Mineshige S., and Wood J.A., 1989, in {\it Theory of Accretion
Disks}, eds. F. Meyer, W. Duschl, J. Frank and E. Meyer-Hofmeister, Kluwer, Dordrecht, p.221

\rea  Moll, R.\ 2009, \aap{507}, 1203 

\rea Nakamura K.E., Matsumoto R., Kusunose M., Kato S., 1996, \pasj{48}, 769

\rea  Narayan R., Goldreich P., Goodman,J., 1987, \mnras{228}, 1

\rea  Narayan R. and Popham R., 1993, in {\it Theory of Accretion Disks II}, eds. W. Duschl et al., Kluwer Dordrecht, p293

\rea Narayan R., Yi I., 1994, \apj{428}, L13

\rea Narayan R., Yi I., Mahadevan R., 1995, {\it Nature} {\bf 374}, 623

\rea Narayan R., Kato S., Honma F., 1997, \apj{476}, 49

\rea  O'Donoghue D.O., Chen A., Marang F., Mittaz P.D., Winkler H., Warner B. 1991, \mnras{250}, 363

\rea Ogilvie G.I., 1999, \mnras{306}, L9

\rea Osaki Y., 1974, \pasj{26}, 429

\rea  Ostriker E.C., Gammie C.F., Stone J.M., 1999, \apj{513}, 259

\rea  Osaki Y., 1993 in {\it Theory of Accretion Disks, II}, eds. W.
Duschl et al., Kluwer Dordrecht, p93

\rea  Paczy\'nski B., 1978, {\it Acta Astron.} {\bf 28}, 91

\rea  Paczy\'nski B., 1991, \apj{370}, 597

\rea  Paczynski, B.\ 1998, \apjl{494}, L45 

\rea  Papaloizou, J.~C.~B., \& Lin, D.~N.~C.\ 1995, \araa, 33, 505 

\rea  Piran T., 1978, \apj{221}, 652

\rea Podolak M., Hubbard W.B. Pollack J.B., 1993, in {\it Protostars and
Planets II}, University of Arizona Press, p.1109

\rea  Popham R. and Narayan R., 1991, \apj{370}, 604

\rea Popham R., Woosley S.E., Fryer C., 1999, \apj{518}, 356

\rea  Popham R., 1997, in Accretion Phenomena and related Outflows (IAU Colloquium 163), ASP Conference Series 121, ed. D. T. Wickramasinghe; G.V.\ Bicknell; and L.\ Ferrario, p.230

\rea  Pringle J.E., 1981 {\it Ann. Rev. Astron. Astrophys.} {\bf 19}, 137

\rea  Pringle J.E. and Savonije, G.J., 1979, \mnras{187}, 777

\rea Quataert E., Narayan R., 1999 \apj{517}, 101

\rea  Rafikov, R.~R.\ 2009, \apj, 704, 281 

\rea  Rappaport S., Podsiadlowski P., 2000, \apj{529}, 946

\rea  Rappaport, S.A., Fregeau, J.M., Spruit, H.C., 2004, \apj{606}, 436

\rea  R\'o\.zyczka M. and Spruit H.C., 1993, \apj{417}, 677 (with video) 

\rea  Rees M.J., 1978, {\it Physica Scripta} {\bf 17}, 193

\rea  Rees M.J., Phinney E.S., Begelman M.C., Blandford R.D., 1982, {\it Nature} {\bf 295}, 17

\rea  Regev, O., Hougerat, A. 1988, \mnras{232}, 81

\rea  Revnivtsev, M., Gilfanov, M., \& Churazov, E.\ 1999, \aap{347}, L23 

\rea  Rincon, F., Ogilvie, G.~I., \& Cossu, C.\ 2007, \aap{463}, 817 

\rea  Rybicki G.R. and Lightman A.P., 1979, {\it Radiative Processes in Astrophysics}, Wiley, New York, Ch 1.5

\rea  Rutten R., van Paradijs J., Tinbergen J., 1992, \aaa{260}, 213

\rea  Sakimoto P., Coroniti F.V., 1989, \apj{342}, 49

\rea  Sawada K., Matsuda T., Inoue M. \& Hachisu 1987, \mnras{224}, 307

\rea  Schandl S., Meyer F., 1994,  \aaa{289}, 149

\rea  Schandl S., Meyer F., 1997, \aaa{321}, 245

\rea Schultz A.L., Price R.H., 1985, \apj{291}, 1

\rea  Shakura N.I. and Sunyaev R.A., 1973, \aaa{24}, 337

\rea  Shi, J., Krolik, J.~H., \& Hirose, S.\ 2010, \apj{708}, 1716 

\rea  Sunyaev R.A., Shakura N.I., 1977, {\it Pi'sma Astron. Zh.} {\bf 3}, 262  (translation {\it Soviet Astron. L.} {\bf 3}, 138)

\rea  Shapiro S.L., Lightman A.P., Eardley D.M., 1976 \apj{204}, 187

\rea  Smak J., 1971, {\it Acta Astron.} {\bf 21}, 15

\rea  Smak J., 1984, \pasp{96}, 54

\rea  Spitzer L., 1965, Physics of fully ionized gases, Interscience Tracts on Physics and Astronomy, New York: Interscience Publication, 1965, 2nd rev. 

\rea  Spruit H.C., 1987, \aaa{184}, 173

\rea  Spruit H.C. 1996, in Evolutionary Processes in Binary Stars, R.A.M.J.
Wijers et al. (eds.), NATO ASI {\bf C 477}, Kluwer Dordrecht, p249

\rea  Spruit H.C., Rutten R.G.M., 1998, \mnras{299}, 768

\rea  Spruit H.C., Matsuda T., Inoue M., Sawada K., 1987, \mnras{229}, 517

\rea  Spruit H.C., Taam R.E., 1993, \apj{402}, 593

\rea  Spruit H.C., 1997, in Accretion disks-new aspects, eds. E.\ 
Meyer-Hofmeister and H.C.\ Spruit, Lecture Notes in Physics {\bf 487}, Springer, p. 67

\rea  Spruit, H.C., \& Deufel, B.\ 2002, \aap{387}, 918 

\rea  Spruit,H.C., 2008, arXiv:0804.3096v4  [astro-ph]

\rea  Spruit, H.C., 2010, arXiv:1004.4545 

\rea  Strom S.E., Edwards S., Skrutskie M.F., 1993, in {\it Protostars and Planets III}, eds. E.H. Levy, J.I. Lunine, Univ. Arizona Press, Tucson, p.837

\rea  Sunyaev R.A., Shakura N.I., 1977, {\it PAZh} {\bf 3}, 262 {it Soviet
Astron. Letters} {\bf 3},138

\rea  Sunyaev, R.~A., \& Titarchuk, L.~G.\ 1980, \aap{86}, 121 

\rea  Taam R.E., 1994, in Compact stars in Binaries (IAU Symp {\bf 165}),
J. van Paradijs et al. eds., Kluwer, p.3

\rea  Taam, R.~E., \& Sandquist, E.~L.\ 2000, \araa{38}, 113 

\rea  Treves A., Maraschi L., and Abramowicz M., 1988 \pasp{100}, 427 

\rea  Turner, N.~J.\ 2004, \apjl{605}, L45 

\rea  Turner, N.~J., Blaes,  O.~M., Socrates, A., Begelman, M.~C., \& Davis, S.~W.\ 2005, \apj{624}, 267 

\rea  van Paradijs J. and McClintock J.E., 1995, in Lewin, W.H.G., van
Paradijs, J. and van den Heuvel, E.P.J., eds., {\it X-ray Binaries}, Cambridge
Univ. Press, Cambridge, p.58

\rea  van Paradijs J, Verbunt F. 1984, in S.E. Woosley, ed., {\it High
Energy Transients in Astrophysics}, AIP Conference Procs. {\bf 115}, p.49

\rea  Velikhov, 1959, {\it J. Expl. Theoret. Phys.} (USSR), {\bf 36}, 1398

\rea Wang J.M., Zhou Y.Y., 1999 \apj{516}, 420

\rea  Warner B. 1987, \mnras{227}, 23

\rea  Warner B. 1995, Cataclysmic Variable Stars, Cambridge: CUP

\rea  Wood J.A., Horne K., Berriman G, Wade R.A. 1989a \apj{341}, 974

\rea  Wood J.A., Marsh T.R., Robinson E.L. et al., 1989b, \mnras{239}, 809

\rea  Wood J.H., Horne K., Vennes S., 1992, \apj{385}, 294

\rea  Woosley, S.~E.\ 1993, \apj{405}, 273 

\rea  Yi I., 1999, in Astrophysical Discs, eds.\ J.A.\ Sellwood and J.\ Goodman, Astronomical Society of the Pacific Conference series {\it 160}, 279. (astro-ph/9905215)

\rea Zdziarski A.A., 1998, \mnras{296}, L51

\rea  Zdziarski, A.~A., Lubinski, P., \& Smith, D.~A.\ 1999, \mnras{303}, L11 

\rea  Zel'dovich Ya. B., 1981, {\it Proc. Roy. Soc. London} {\bf A374}, 299

\end{document}